\documentclass{article}
\input epsf

\begin{document}

\makeatletter
\def\@maketitle{\newpage
 \null
 {\normalsize \tt \begin{flushright} 
  \begin{tabular}[t]{l} \@date 
  \end{tabular}
 \end{flushright}}
 \begin{center}
 \vskip 2em
 {\LARGE \@title \par} \vskip 1.5em {\large \lineskip .5em
 \begin{tabular}[t]{c}\@author 
 \end{tabular}\par} 
 \end{center}
 \par
 \vskip 1.5em} 
\makeatother
\topmargin=-1cm
\oddsidemargin=1.5cm
\evensidemargin=-.0cm
\textwidth=15.5cm
\textheight=22cm
\setlength{\baselineskip}{16pt}
\title{ An Associative and Noncommutative Product 
for the Low Energy Effective Theory of a D-Brane in
Curved Backgrounds and  Bi-Local Fields}
\author{  Kiyoshi {\sc Hayasaka}\thanks{ hayasaka@particle.sci.hokudai.ac.jp}
        \  and Ryuichi {\sc Nakayama}\thanks{
                          nakayama@particle.sci.hokudai.ac.jp} 
\\[1cm]
{\small
    Division of Physics, Graduate School of Science,} \\
{\small
           Hokkaido University, Sapporo 060-0810, Japan}
}
\date{
  EPHOU-01-004  \\
hep-th/0109125 \\ 
September 2001  
}
%
%
\maketitle

\begin{abstract}

We point out that when a D-brane is placed in an NS-NS $B$ field background 
with non-vanishing field strength ($H=dB$) along the D-brane worldvolume, the 
coordinate of one end  of the open string does not commute with that of the 
other in the low energy limit. The degrees of the freedom associated with 
both ends are not decoupled and accordingly, the effective action must be 
quite different from that of the ordinary noncommutative gauge theory for a 
constant $B$  background.  We construct an associative and noncommutative 
product {\boldmath $\star$} which operates on the coordinates of both ends of 
the string and propose a new type of noncommutative gauge  action for the 
low energy effective theory of a D$p$-brane. This effective theory is bi-local 
and lives in twice as large dimensions ($2D=2(p+1)$) as in the $H=0$ case.  
When viewed as a theory in the $D$-dimensional space, this theory is 
non-local and we must force the two ends of the string to coincide. We will 
then propose a prescription for reducing this bi-local effective action to 
that in $D$ dimensions and obtaining a local effective action. 

\end{abstract}
\newpage
\setlength{\baselineskip}{18pt}

\newcommand {\beq}{\begin{equation}}
\newcommand {\eeq}{\end{equation}}
\newcommand {\beqa}{\begin{eqnarray}}
\newcommand {\eeqa} {\end{eqnarray}}
\newcommand{\bm}[1]{\mbox{\boldmath $#1$}}
\newcommand{\al}{2\pi \alpha'}

\section{Introduction}
\hspace{5mm}
Noncommutative field theories\cite{CDS}\cite{DH} are obtained in an $\alpha' 
\rightarrow 0$ limit of the open string theory. This noncommutativity is 
inherent in the open string theory\cite{W}. By considering an open string in 
an NS-NS $B$ field background and taking a specific field theory limit the 
effects of the $B$ field are encoded into a special multiplication rule of 
functions, the $*$ product or Moyal product.  By studying such field theories 
we expect that open string physics may be understood in the framework of 
field theory. Until recently such investigations are mostly restricted to 
constant $B$ backgrounds\cite{Schomerus}\cite{Open}\cite{Ho1}\cite{SW}. The 
algebra of the coordinates of the end points of the open string was studied, 
the connection of the commutative and noncommutative descriptions of the 
gauge theory was elucidated, and so on.

When the $B$ field is constant, the background metric $g_{\mu \nu}$ is flat.
Attempts to extend the analysis to the curved background, {\em i.e.}, the 
nonvanishing $B$ field strength $H=dB$, appeared in the context of open 
string theory in WZW models\cite{fuzzysphere}.  In this  approach it was 
found that
the algebra of functions on the D-brane worldvolume is given by the 
$q$-deformation of the Lie algebra.  The general framework for general 
backgrounds was, however, not yet obtained.

More recently a noncommutative field theory for a weak field strength $H$
and a weakly curved background was investigated\cite{CS}.  
\footnote{Topological sigma models with a $H=dB$ term 
were also considered in \cite{Hpoisson}  and the related underlying geometry 
was called H-Poisson or twisted Poisson one.}
Several correlation functions of the string theory are obtained in the 
$\alpha' \rightarrow 0$ limit and it was concluded that the algebra of 
functions on the D-brane worldvolume is non-associative and noncommutative. 
Various correlation functions are expressed in terms of a non-associative 
product $\bullet$.
In \cite{Ho2} the commutation relations of the coordinate $x$ of the end point 
of the open string were studied by using the approximation of a very short 
and slowly moving string. It was found that the commutators of the 
coordinate  $x$ contain a momentum $p$ and it was argued that the space on 
which the D-brane lives is not the usual noncommutative space which we are 
getting used to.  In \cite{Ho3} a new product $\diamond$ was defined in order
to make the product $\bullet$ associative on the functions of $x$ and $p$ and 
an attempt to define a gauge transformation in this `$x$-$p$ space' 
was presented. An explicit construction of the gauge theory action, however,
was not obtained yet.  

The purpose of the present paper is to understand how the low energy 
effective theory can be formulated as an associative and noncommutative gauge 
theory, when the background space is curved and the field strength $H=dB$ 
along the direction of the D-brane worldvolume is nonvanishing. Our analysis 
will be restricted to the bosonic string. We assume that the field strength 
is weak and perform analysis in perturbation series in $H$ up to ${\cal O}
(H^1)$. We will first obtain the commutation relations of the coordinates 
$x^{\mu}$, $y^{\mu}$ of the two ends (at $\sigma=0$ and $\pi$) of the open 
string in the low energy limit.(sec 2) This will be performed by discarding 
the oscillators of the string variable $X^{\mu}(\tau,\sigma)$, but without 
further approximation.  The result is striking.  $x^{\mu}$ and $y^{\mu}$ do 
not commute with each other in contrast to the $H=0$ case. This means that 
when we take the $\alpha' \rightarrow 0$ limit, the gauge and scalar degrees 
of freedom at both ends of the string are not decoupled in the low energy 
effective theory.  We are forced to take into account the degrees of freedom 
at both ends in constructing the field theory description and the numbers of 
gauge and scalar fields are doubled. Furthermore the noncommutative product 
{\boldmath $\star$} constructed according to the algebra of $x$ and $y$ is 
{\em associative}, but operates on both coordinates.(sec 3) We are thus lead 
to consider a bi-local field theory. All fields are functions of $x$ and $y$. 
The gauge theory must be formulated in the 2$D$ dimensional $(x,y)$ space 
instead of the ordinary $D$ dimensional $x$ space. This bi-local field theory 
will, however, be acausal and must be regarded as an intermediate step toward 
construction of the local effective field theory, which will be discussed later. 

Construction of this bi-local  gauge theory can be performed in a standard 
way. The derivatives are generalized in such a way that they  satisfy 
Leibnitz rule with respect to the product {\boldmath $\star$}.(sec 4)  The 
commutator of the gauge covariant derivatives defined in terms of these new 
derivatives gives the gauge field strength.(sec 5) The trace or the integral 
is defined in such a way that the cyclic property is respected.
An obstruction in this prescription is the curved D-brane worldvolume.
The metric of the effective gauge theory is the open string metric\cite{SW},
which is curved already at ${\cal O}(H^1)$\cite{CS}.  Because the gauge 
symmetry of the noncommutative gauge theory is realized by differential 
operators on the fields, if the action contains such a non-constant metric, 
one anticipates that the gauge symmetry may be broken.  Remarkably, we found 
that there exists a coordinate transformation in the $(x^{\mu}, y^{\mu})$ 
space (not in the $x^{\mu}$ space) which effectively makes $H=0$. In other 
words in a suitably chosen frame $(x^{\prime \mu}(x,y), y^{\prime \mu}(x,y))$ 
the coordinates satisfy the commutation relations of the $H=0$ theory.  
Therefore it is natural to further assume that the metric is flat in this 
frame.  Moreover the derivatives with respect to $x'$ and $y'$ 
coincide with the new derivatives defined above to satisfy Leibnitz rule.
By writing down the action for the noncommutative gauge theory with
$H=dB=0$ in the flat $(x',y')$ frame and by going to the $(x,y)$ frame by 
means of the  inverse coordinate transformation, we obtain the action for a
bi-local, noncommutative gauge theory in curved backgrounds.

As mentioned above, the above bi-local theory is non-local and will be acausal.
Next we will present a prescription to reduce our bi-local theory to a 
`local theory'.(sec 6) Actually, in order to obtain a 
massless string we  must consider a limit $y \rightarrow x$. This is not 
automatically achieved in the $\alpha' \rightarrow 0$ limit. We will set 
$y=x$ by hand in the integrand of the effective action and integrate this 
over $x$ with a suitable weight function $\omega(x)$. 
We will show cyclic property of this reduced effective action which relates 
various correlation functions of the fields at $x$ and $y$, {\it i.e.}, at
both ends of the string. This reduced effective action is still invariant 
under a part of the gauge transformation. 
We propose this as the low energy effective action of the open string 
in the $H \neq 0$ background. This reduced effective action, however, contains
{\boldmath $\star$} in the integrand and the manipulation such as 
differentiation with respect to $x$ and/or $y$ must be performed before 
setting $y=x$.  In this sense this reduced effective action is not the 
ordinary action in the $x$ space. 

Finally in sec 7 we will speculate on the relation of our noncommutative 
gauge theory to the Matrix model in curved backgrounds.

\section{Noncommutativity of the two end-point coordinates of an open string}
\hspace{5mm}
We will quantize an open string theory with a D$p$-brane in the presence of an 
NS-NS $B$ field along the D$p$-brane, whose worldvolume spans a subspace
$\mu=0,1,2, \ldots,p$ .  The field strength of the $B$ field, $H=dB$, 
is assumed to be small but non-vanishing in the direction of the D$p$-brane 
worldvolume. 

In this case we can perform a perturbative analysis in $H$.  Such an analysis 
was first attempted in \cite{CS}.  We will simplify the problem by taking 
$H_{\mu \nu \lambda}$ constant. By the consistency of the background due to 
Weyl invarinace the metric tensor $g_{\mu \nu}$ must be a constant up to 
${\cal O}(H^1)$ \cite{CS}. We will  choose a gauge 
\begin{equation}
B_{\mu \nu}(X)= b_{\mu \nu} + \frac{1}{3} H_{\mu \nu \lambda} X^{\lambda},  
\end{equation}
where $b_{\mu \nu}$ is a constant.  We assume that $H_{\mu \nu \lambda}$ is
nonvanishing for $\mu, \nu, \ldots=1,2,\ldots,D (\leq p)$.  
When $ H_{\mu \nu \lambda} \neq 0$, $b_{\mu \nu}$ is also in general  
non-vanishing because we can always shift $X^{\mu}$ by a constant.
We also assume that $B$ is block diagonal, {\em i.e.}, $B_{\mu i}=B_{i \mu}=0
\ $ for $\mu \leq D, \ i > D$ and $D$ is even and 
$\det (b_{\mu \nu})_{\mu, \nu=1,2, \ldots,D} \neq 0$.  
In what follows we will restrict our attention only to this $D$-dimensional 
subspace. 

The worldsheet action of the open string in the conformal gauge is given by 
\begin{equation}
S = \frac{1}{4\pi \alpha'} \int d\tau d\sigma \{ g_{\mu \nu} \partial_{\alpha}
X^{\mu} \partial^{\alpha} X^{\nu} + B_{\mu \nu} \epsilon^{\alpha \beta}
\partial_{\alpha} X^{\mu} \partial_{\beta} X^{\nu} \}.
\end{equation}
The coordinate $X^{\mu}$ must satisfy the following boundary conditions
at $\sigma = 0, \pi$. 
\begin{equation}
g_{\mu \nu} \partial_{\sigma} X^{\nu} - B_{\mu \nu} \partial_{\tau} X^{\nu}=0
\label{BC1}
\end{equation}
Inside the worldsheet it must satisfy the equations of motion.
\begin{equation}
g_{\lambda \rho} (\partial^{\alpha}\partial_{\alpha} X^{\rho}+
{\Gamma_{\mu \nu}}^{\rho} \partial_{\alpha}X^{\mu} \partial^{\alpha}X^{\nu})
- H_{\mu \nu \lambda} \partial_{\tau} X^{\mu} \partial_{\sigma}X^{\nu}=0
\label{EoM1}
\end{equation}

For small $H$ the metric $g$ is constant up to ${\cal O}(H^1)$ and eq of 
motion (\ref{EoM1}) reads
\begin{equation}
- \ddot{X}^{\mu} + X^{\prime \prime \mu} -g^{\mu \nu} H_{\nu \lambda \rho} 
\dot{X}^{\lambda} X^{\prime \rho}=0 + {\cal O}(H^2),
\label{EoM2}
\end{equation}
where the dot and prime stand for $\partial / \partial \tau$ and 
$\partial / \partial \sigma$, respectively.  
The boundary condition (\ref{BC1}) is now expressed as
\begin{equation}
X^{\prime \mu} - {J^{\mu}}_{\nu} \dot{X}^{\mu} - \frac{1}{3}g^{\mu \nu}
H_{\nu \lambda \rho}\dot{X}^{\lambda}X^{\rho}=0 \qquad (\mbox{at} \quad
\sigma=0, \pi),
\label{BC2}
\end{equation}
where ${J^{\mu}}_{\nu}=g^{\mu \lambda}b_{\lambda \nu}$.

We will solve (\ref{EoM2}) under the condition (\ref{BC2}) up to 
${\cal O}(H^1)$.  We set $X^{\mu}=X^{(0) \mu}+X^{(1) \mu}$.
The leading order solution is given by
\begin{equation}
X^{(0)\mu}(\tau, \sigma)= x^{\mu}+ \tau p^{\mu}+\sigma (Jp)^{\mu}
+ {(1+J)^{\mu}}_{\nu} f^{\nu}(\tau+\sigma)+ {(1-J)^{\mu}}_{\nu}
f^{\nu}(\tau-\sigma),
\label{X0}
\end{equation}
where $(Jp)^{\mu}$ stands for ${J^{\mu}}_{\nu}p^{\nu}$ and 
\begin{equation}
f^{\mu}(\sigma)= \sum_{n=1}^{\infty} \frac{1}{n} (a^{\mu}_n e^{-in \sigma}
+ a^{\mu}_{-n} e^{in \sigma}).
\end{equation}
$x^{\mu}$, $p^{\mu}$ are `zero modes' and $a_n^{\mu}$ oscillators.
We will next plug this in (\ref{EoM2}) and derive an equation for the 
${\cal O}(H^1)$ correction $X^{(1) \mu}$. This can easily be solved.
\begin{eqnarray}
X^{(1) \mu} &=& -\frac{1}{4}(\tau^2-\sigma^2) \ g^{\mu \nu} \ H_{\nu \lambda 
\rho} \ p^{\lambda} \ (Jp)^{\rho} -\frac{1}{4} (\tau-\sigma) \ g^{\mu \nu} \ 
H_{\nu \lambda \rho} \ p^{\lambda} {(1+J)^{\rho}}_{\alpha } \ 
f^{\alpha}(\tau+\sigma) \nonumber \\
&&+\frac{1}{4} (\tau+\sigma) \  g^{\mu \nu} \ H_{\nu \lambda \rho} \ 
p^{\lambda} \ {(1-J)^{\rho}}_{\alpha} \ f^{\alpha}(\tau-\sigma) \nonumber \\
&&+\frac{1}{4} (\tau-\sigma) \ g^{\mu \nu} \ H_{\nu \lambda \rho} \ 
(Jp)^{\lambda} \ {(1+J)^{\rho}}_{\alpha} \ f^{\alpha}(\tau+\sigma) \nonumber \\
&&+\frac{1}{4} (\tau+\sigma) \ g^{\mu \nu} \ H_{\nu \lambda \rho} \ 
(Jp)^{\lambda} \ {(1-J)^{\rho}}_{\alpha} \ f^{\alpha}(\tau-\sigma) \nonumber \\
&&+\frac{1}{2}g^{\mu \nu} \ H_{\nu \lambda \rho} \ {(1+J)^{\lambda}}_{\alpha}
\ {(1-J)^{\rho}}_{\beta} \ f^{\alpha}(\tau+\sigma) \ f^{\beta}(\tau-\sigma) 
\nonumber \\
&&+ g^{\mu \nu} \ h_{\nu} \ (\tau+\sigma)+g^{\mu \nu} \ k_{\nu} \ 
(\tau-\sigma)
\label{X1}
\end{eqnarray}
Here $h_{\nu}$ and $k_{\nu}$ are some functions to be determined by 
(\ref{BC2}).

The energies of the oscillation modes are proportional to $1/\sqrt{\alpha'}$ 
and in the $\alpha' \rightarrow 0$ limit they can be ignored.  
In this paper we will restrict attention only to the `zero modes' and simply
set $a^{\mu}_n=0 \ \ (f^{\mu}=0)$. 
One may speculate that $x,p$ and $a$ may mix up due to the interaction and
the commutators of $x,p$ may contain $a$. We have explicitly checked that 
this is not the case up to ${\cal O}(H^1)$. The full analysis including 
the oscillators 
is now in progress and will be reported elsewhere\cite{HNS}. 
Eq (\ref{BC2}) forces $h_{\nu}$, $k_{\nu}$ to satisfy an equation
\begin{eqnarray}
&&{(1+J)^{\nu}}_{\mu} h'_{\nu}(\tau+\sigma)-{(1-J)^{\nu}}_{\mu} 
k'_{\nu}(\tau-\sigma) \nonumber \\
&&\qquad \qquad = -\frac{1}{3}H_{\mu \nu \lambda}x^{\nu}p^{\lambda}-
\frac{1}{6}\sigma H_{\mu \nu \lambda} (Jp)^{\lambda}p^{\nu} 
+\frac{1}{2} \tau {J^{\nu}}_{\mu}H_{\nu \lambda \rho} p^{\lambda}(Jp)^{\rho}
\label{BC3}
\end{eqnarray}
at $\sigma =0, \pi$.  The primes on $h_{\nu}$, $k_{\nu}$ stand for the 
derivatives with respect to the arguments. This equation can be solved by 
subtracting (\ref{BC3}) with
$\sigma=0$ from (\ref{BC3}) with $\sigma=\pi$ and $\tau \rightarrow \tau+\pi$.
This determines $h_{\mu}$ and then (\ref{BC3}) gives $k_{\mu}$.
\begin{eqnarray}
h_{\mu}(\tau) &=& \frac{\tau^2}{24} {\left( \frac{1-3J}{1+J} 
\right)^{\nu}}_{\mu} \  H_{\nu \lambda \rho}(Jp)^{\lambda}p^{\rho}
-\frac{\tau}{6} {\left( \frac{1}{1+J} \right)^{\nu}}_{\mu} \  H_{\nu \lambda 
\rho} x^{\lambda} p^{\rho} \ \ , \nonumber \\
k_{\mu}(\tau) &=& \frac{\tau^2}{24} {\left( \frac{1+3J}{1-J} 
\right)^{\nu}}_{\mu} \ H_{\nu \lambda \rho}(Jp)^{\lambda}p^{\rho}
+\frac{\tau}{6} {\left( \frac{1}{1-J} \right)^{\nu}}_{\mu} \ H_{\nu \lambda 
\rho} x^{\lambda} p^{\rho} .
\end{eqnarray}

The next step is to quantize this system.  We will use the phase-space 
integral invariant of Poincar\'e $I$ to derive Poisson brackets\cite{Ho1}. 
It is defined by 
\begin{equation}
I = \int_0^{\pi} d\sigma \delta X^{\mu}(\tau, \sigma) \wedge \delta P_{\mu}
(\tau, \sigma),
\label{invariant}
\end{equation}
where $P_{\mu}$ is the canonical momentum conjugate to $X^{\mu}$,
\begin{equation}
2\pi \alpha' P_{\mu}= -g_{\mu \nu} \dot{X}^{\nu} + B_{\mu \nu} X^{\prime \nu}.
\label{P}
\end{equation} 
$\delta X^{\mu}$ and  $\delta P_{\mu}$ represent differential 1-forms.
The quantity (\ref{invariant}) is indepedent of $\tau$ and defines the 
Poisson structure. Substitution of (\ref{X0}), (\ref{X1}), (\ref{P}) into 
(\ref{invariant}) yields $I$ as 
a 2-form in the basis $\delta x^{\mu} \wedge \delta x^{\nu}$, $\delta x^{\mu} 
\wedge \delta p^{\nu}$, $\delta p^{\mu} \wedge \delta p^{\nu}$.  
The coefficients are a 2$D$  by 2$D$  matrix and its inverse gives the Poisson 
brackets. They are then converted to the commutation relations.
\begin{eqnarray}
 \ [x^{\mu}, x^{\nu} ] &=& 2\pi \alpha' \{  i \theta^{\mu \nu} + 
\frac{i}{3} (\theta^{\mu \lambda} \theta^{\nu \rho} - G^{\mu \lambda} 
G^{\nu \rho} ) H_{\lambda \rho \sigma} \  
x^{\sigma} \nonumber \\
&&+ \frac{\pi i}{9} (3\theta^{\mu \lambda} \theta^{\nu \rho} -
G^{\mu \lambda} 
G^{\nu \rho} ) \ H_{\lambda \rho \sigma} \ (Jp)^{\sigma} 
 - \frac{\pi i}{9}(\theta^{\mu \lambda} G^{\nu \rho}+G^{\mu \lambda} 
\theta^{\nu \rho}) \ H_{\lambda \rho \sigma} \  p^{\sigma} \}, \nonumber \\
\ [ x^{\mu}, p^{\nu}] &=& 2\pi \alpha' \{ -\frac{i}{\pi}  G^{\mu \nu}+\frac{i}
{3\pi} \theta^{\mu \lambda} G^{\nu \rho} \ H_{\lambda \rho \sigma} \  
x^{\sigma} \nonumber \\
&& +\frac{i}{6}(3\theta^{\mu \lambda} \theta^{\nu \rho} -
G^{\mu \lambda} 
G^{\nu \rho} ) \ H_{\lambda \rho \sigma} \ p^{\sigma} 
-\frac{i}{2}(G^{\mu \lambda} \theta^{\nu \rho}-\theta^{\mu \lambda} 
G^{\nu \rho}) \  H_{\lambda \rho \sigma} \  (Jp)^{\sigma} \}, \nonumber \\
\ [ p^{\mu}, p^{\nu}] &=&  \frac{4i}{3}\alpha' \ G^{\mu \lambda}
G^{\nu \rho} \ H_{\lambda \rho \sigma} (Jp)^{\sigma}.
\label{CRpp}
\end{eqnarray}
Here $\theta$ and $G$ are defined by
\begin{equation}
\theta^{\mu \nu} = \left( \frac{-J}{1-J^2} g^{-1} 
\right)^{\mu \nu}, \qquad
G^{\mu \nu} =  \left( \frac{1}{1-J^2} g^{-1} \right)^{\mu \nu}.
\label{thetaG}
\end{equation}
These are the noncommutative parameter\footnote{This $\theta$ is different 
from that used in the literature by a factor $2\pi \alpha'$. When multiplied 
by this factor, $\theta$ remains finite in the $\alpha' \rightarrow 0$ limit. 
We used the definition (\ref{thetaG}) in order to avoid messy expression for 
(\ref{CRpp}).} and the open string metric for the $H=0$ background\cite{SW}.

Similar commutation relations in the $\alpha' \rightarrow 0$ limit are 
obtained in \cite{Ho2} by using some approximation of short and slowly moving 
string. This $\alpha' \rightarrow 0$ limit is a specific one taken with fine 
tuning\cite{SW}.  
$g_{\mu \nu}$ is adjusted to be ${\cal O}(\alpha^{\prime 2})$, while 
$ \ x^{\mu} \sim {\cal O}(\alpha^{\prime 0})$, $ \ b_{\mu \nu}, H_{\mu \nu 
\lambda}, p^{\mu} \sim {\cal O}(\alpha^{\prime 1})$. Hence 
${J^{\mu}}_{\nu} \sim {\cal O}(\alpha^{\prime -1})$, $\theta^{\mu 
\nu} \sim  {\cal O}(\alpha^{\prime -1})$ and $ \ G^{\mu \nu} \sim  {\cal O}
(\alpha^{\prime 0})$. 
Defining the total momentum at $\tau=0$ by $\tilde{p}_{\mu} \equiv 
\int_0^{\pi} d\sigma P_{\mu}(0,\sigma) \sim {\cal O}(\alpha^{\prime 0})$, 
we get in the $\alpha' \rightarrow 0$ limit
\begin{eqnarray}
\ [ x^{\mu}, x^{\nu} ]& =& 2\pi \alpha' \left(i \hat{\theta}^{\mu \nu}(x) 
+\frac{2\pi i}{3} \ \alpha'\theta^{\mu \lambda}\theta^{\nu \rho}  \ H_{\lambda 
\rho \sigma} \ \theta^{\sigma \alpha} \ \tilde{p}_{\alpha} \right), 
\nonumber \\
\ [x^{\mu}, \tilde{p}_{\nu} ] &=& i \delta^{\mu}_{\nu} + 
\frac{\pi i}{3} \alpha' \theta^{\mu \lambda} \theta^{\rho \alpha} 
\ H_{\nu \lambda \rho} \ \tilde{p}_{\alpha}, \qquad 
[ \tilde{p}_{\mu}, \tilde{p}_{\nu}] = 0,  
\end{eqnarray}
which agrees with the result of \cite{Ho2}, \cite{Ho3}. Here 
$\hat{\theta} (x)$ is $\theta$ in (\ref{thetaG}) with $J=g^{-1} b$ replaced 
by $g^{-1}B(x)$.   

In \cite{Ho2} it is argued that the appearence
of $p^{\mu}$ on the righthand sides of (\ref{CRpp}) makes the construction of 
the low energy effective action difficult.  In \cite{Ho3} some proposal for 
a construction of gauge transformations is presented.  An explicit form of 
the action integral in the $x$-$p$ space is, however, not proposed.

In this paper we will observe the relations (\ref{CRpp}) from a
different view point and construct a noncommutative gauge theory based on this 
algebra. Crucial point is that the coordinates of the end points of the 
open string (at $\tau=0$) are given by
\begin{eqnarray}
\left. X^{\mu}(0,0) \right|_{\rm{zero} \ \rm{ modes}} &=& x^{\mu}, \nonumber \\
\left. X^{\mu}(0,\pi) \right|_{\rm{zero} \ \rm{ modes}} &=& x^{\mu}+
\pi (Jp)^{\mu} +{\cal O}(H^1).
\end{eqnarray}
Throughout this paper we will use $x$ and $y$ to stand for the coordinates
of the two ends of the open string.
The difference of the two end point coordinates is thus proportional to $p$.
\begin{equation}
y^{\mu}-x^{\mu} = \pi (Jp)^{\mu} + {\cal O}(H^1)
\label{yxp}
\end{equation}
Note that this is ${\cal O}(\alpha^{\prime 0})$ and does not automatically
go to zero but is fixed in the $\alpha' \rightarrow 0$ limit. For large
momenta the length of the open string grows.\cite{SST} If $b_{\mu \nu}$ is 
invertible, which we assume in this paper, so are ${J^{\mu}}_{\nu}$ and 
$\theta^{\mu \nu}$, and we can invert (\ref{yxp}).
\begin{equation}
p^{\mu}=\frac{1}{\pi} {(J^{-1})^{\mu}}_{\nu} \ (y^{\nu}-x^{\nu}) + 
{\cal O}(H^1) 
\label{pyx}
\end{equation}
Now the algebra (\ref{CRpp}) can be reexpressed in terms of 
$x^{\mu}$ and $y^{\mu}$.\footnote{In what follows we will set 
$2\pi \alpha'=1$.}
\begin{eqnarray}
\ [ x^{\mu}, x^{\nu} ] & = &  i \theta^{\mu \nu} 
+ \frac{i}{3} (\theta^{\mu \lambda}\theta^{\nu \rho} -
G^{\mu \lambda}G^{\nu \rho}
) H_{\lambda \rho \alpha} x^{\alpha} 
 + \frac{i}{9} (3\theta^{\mu \lambda}\theta^{\nu \rho} 
- G^{\mu \lambda}G^{\nu \rho}
) H_{\lambda \rho \alpha}(y^{\alpha} - x^{\alpha}) \nonumber \\
& & \qquad - \frac{i}{9} (\theta^{\mu \lambda}G^{\nu \rho} 
+G^{\mu \lambda}\theta^{\nu \rho}) H_{\lambda \rho \alpha}
{(J^{-1})^{\alpha}}_{\beta}(y^{\beta} - x^{\beta}), 
\label{CR1}
\end{eqnarray}

\begin{eqnarray}
\ [ x^{\mu}, y^{\nu} ] & = & 
-\frac{i}{18}  (3\theta^{\mu \lambda}\theta^{\nu \rho} -
G^{\mu \lambda}G^{\nu \rho}
)H_{\lambda \rho \alpha}(y^{\alpha} - x^{\alpha}) \nonumber \\
& & + \frac{i}{18} (\theta^{\mu \lambda}G^{\nu \rho} +
G^{\mu \lambda}\theta^{\nu \rho}
)H_{\lambda \rho \alpha}{(J^{-1})^{\alpha}}_{\beta}(y^{\beta} - x^{\beta}), 
\label{CR2}
\end{eqnarray}

\begin{eqnarray}
\ [ y^{\mu}, y^{\nu} ] & = & - i  \theta^{\mu \nu}
- \frac{i}{3} (\theta^{\mu \lambda}\theta^{\nu \rho} -
G^{\mu \lambda}G^{\nu \rho} ) H_{\lambda \rho \alpha} y^{\alpha} 
 + \frac{i}{9} (3\theta^{\mu \lambda}\theta^{\nu \rho} 
- G^{\mu \lambda}G^{\nu \rho} ) H_{\lambda \rho \alpha}(y^{\alpha} - 
x^{\alpha}) \nonumber \\
& & \qquad - \frac{i}{9} (\theta^{\mu \lambda}G^{\nu \rho} 
+G^{\mu \lambda}\theta^{\nu \rho}) H_{\lambda \rho \alpha}
{(J^{-1})^{\alpha}}_{\beta}(y^{\beta} - x^{\beta}) 
\label{CR3}
\end{eqnarray}

(\ref{CR1}) and (\ref{CR3}) shows that the coordinates of the ends of the 
open string $x^{\mu}$, $y^{\mu}$ do not commute.  The commutator of $y$'s  
is obtained from that of $x$'s by interchanging $x \leftrightarrow y$ and 
reversing the overall sign. Furthermore (\ref{CR2}) shows that the coordinate 
of one end $x$ does not commute with that of the other $y$. This is in 
contrast to the $H=0$ case\cite{Open}\cite{Ho3}. The gauge and scalar degrees 
of freedom at the $\sigma=0$ end and those at $\sigma=\pi$ are not 
decoupled.  To construct a low energy effective theory we need to consider 
both degrees of freedom. This means that if the two ends of the string are 
separated, the low energy effective field theory will become acausal. 
To avoid this problem we will take a limit $y \rightarrow x$ in sec 5.
Neverthelss these two ends must be separated for a while in order to 
construct an associative product.
 
We expect that the full string theory including the oscillators is  causal.  
The reason for the non-commutativity of $x$ and $y$ may lie in the fact that 
we discarded oscillators. It will be interesting to study the commutators of 
the full string variable $X^{\mu}(\tau, \sigma)$.

\section{An associative and noncommutative product}
\hspace{5mm}
It can be shown that the commutation relations (\ref{CR1})-(\ref{CR3}) satisfy
Jacobi identities. We will then construct a product {\boldmath $\star$} which 
realizes this algebra. The parameter $H_{\mu \nu \lambda}$ is treated as a 
small purturbation but $\theta^{\mu \nu}$ must be taken into account to the 
full order. We report only the results here.
\begin{eqnarray}
&& f_1(x,y) \bm{\star} f_2(x,y) \nonumber \\
& \equiv & f_1(x,y) * f_2(x,y)  \nonumber \\
& & - \frac{i}{6}\left( G^{\mu \lambda} 
G^{\nu \rho}-\theta^{\mu \lambda} 
\theta^{\nu \rho} \right)  H_{\lambda \rho \alpha}
\left( x^{\alpha} \frac{\partial f_1}{\partial x^{\mu}} * 
\frac{\partial f_2}{\partial x^{\nu}} - 
 y^{\alpha} \frac{\partial f_1}{\partial y^{\mu}} *
\frac{\partial f_2}{\partial y^{\nu}} \right) \nonumber \\
& & - \frac{i}{18}\left( G^{\mu \lambda} G^{\nu \rho}-3\theta^{\mu \lambda} 
\theta^{\nu \rho} \right) H_{\lambda \rho \alpha} \nonumber \\
& & \qquad \qquad \times (y^{\alpha}-x^{\alpha}) 
\left( \frac{\partial f_1}{\partial x^{\mu}} * 
\frac{\partial f_2}{\partial x^{\nu}} 
-\frac{1}{2} \frac{\partial f_1}{\partial x^{\mu}} * 
\frac{\partial f_2}{\partial y^{\nu}} 
-\frac{1}{2} \frac{\partial f_1}{\partial y^{\mu}} * 
\frac{\partial f_2}{\partial x^{\nu}} 
+ \frac{\partial f_1}{\partial y^{\mu}} *
\frac{\partial f_2}{\partial y^{\nu}} \right) \nonumber \\
& & +\frac{i}{18}\left( \theta^{\mu \lambda} G^{\nu \rho}+G^{\mu \lambda} 
\theta^{\nu \rho} \right) H_{\lambda \rho \alpha} 
{\left(G^{-1} \theta^{-1} \right)^{\alpha}}_{\beta} \nonumber \\
& & \qquad \qquad \times (y^{\beta}-x^{\beta}) 
\left( \frac{\partial f_1}{\partial x^{\mu}} * 
\frac{\partial f_2}{\partial x^{\nu}} 
-\frac{1}{2} \frac{\partial f_1}{\partial x^{\mu}} * 
\frac{\partial f_2}{\partial y^{\nu}} 
-\frac{1}{2} \frac{\partial f_1}{\partial y^{\mu}} * 
\frac{\partial f_2}{\partial x^{\nu}} 
+ \frac{\partial f_1}{\partial y^{\mu}} *
\frac{\partial f_2}{\partial y^{\nu}} \right) \nonumber \\
&& -\frac{1}{36}G^{\mu \lambda} G^{\nu \rho} \theta^{\beta \alpha} H_{\lambda
\rho \alpha} \nonumber \\
& & \qquad \times \left( \frac{\partial f_1}{\partial x^{\mu}} * 
\frac{\partial^2
 f_2}{\partial x^{\nu} \partial x^{\beta}}
 - \frac{\partial^2 f_1}{\partial x^{\mu} \partial x^{\beta}} * 
\frac{\partial f_2}{\partial x^{\nu}} 
 + \frac{\partial f_1}{\partial y^{\mu}} * 
\frac{\partial^2
 f_2}{\partial y^{\nu} \partial y^{\beta}}
 - \frac{\partial^2 f_1}{\partial y^{\mu} \partial y^{\beta}} * 
\frac{\partial f_2}{\partial y^{\nu}}  \right) \nonumber \\
&& +\frac{1}{72} \left( -3\theta^{\mu \lambda}\theta^{\nu \rho} 
\theta^{\beta \alpha} + G^{\mu \lambda}G^{\nu \rho} \theta^{\beta \alpha}
+ \theta^{\mu \lambda} G^{\nu \rho} G^{\beta \alpha} +
G^{\mu \lambda} \theta^{\nu \rho} G^{\beta \alpha} \right) H_{\lambda \rho 
\alpha} \nonumber \\
& & \qquad \times \left( \frac{\partial f_1}{\partial x^{\mu}} * 
\frac{\partial^2
 f_2}{\partial x^{\nu} \partial y^{\beta}}
 - \frac{\partial^2 f_1}{\partial x^{\mu} \partial y^{\beta} } * 
\frac{\partial f_2}{\partial x^{\nu}} 
  + \frac{\partial f_1}{\partial y^{\mu}} * 
\frac{\partial^2
 f_2}{\partial y^{\nu} \partial x^{\beta}}
 - \frac{\partial^2 f_1}{\partial y^{\mu} \partial x^{\beta} } * 
\frac{\partial f_2}{\partial y^{\nu}}  \right) \nonumber \\
&& 
\label{newstar}
\end{eqnarray}
Here $*$ is the ordinary noncommutative product for $x$ and $y$ in the $H=0$
 case.
\begin{eqnarray}
&&f_1(x,y) * f_2(x,y) \nonumber \\
&\equiv & \left. \exp \left( \frac{i}{2} \theta^{\mu \nu} 
\frac{\partial}{\partial x^{\mu}} \frac{\partial}{\partial x^{\prime \nu}}-
 \frac{i}{2}\theta^{\mu \nu} \frac{\partial}{\partial 
y^{\mu}} \frac{\partial}{\partial y^{\prime \nu}} \right) f_1(x,y)f_2(x',y')
\right|_{x'=x, y'=y}
\label{oldstar}
\end{eqnarray}
The noncommutative parameter is $\theta^{\mu \nu}$ for $x$ and 
$-\theta^{\mu \nu}$ for $y$.

The product (\ref{newstar}) is associative.
The proof of the associativity (up to ${\cal O}(H^1)$)
\begin{equation}
\left(f_1(x,y) \bm{\star} f_2(x,y)\right) \bm{\star} f_3(x,y) 
= f_1(x,y) \bm{\star} \left(f_2(x,y) \bm{\star}
f_3(x,y)\right)
\label{associativity}
\end{equation}
is straightforward. 

Because the product {\boldmath $\star$} contains both  $x$ and $y$, the 
product of the functions of $x$ only, $f_1(x) \bm{\star} f_2(x)$ will also 
depend on $y$. Hence we must regard all fields to be functions of both $x$ 
and $y$, {\em i.e.}, bi-local fields. 

\section{Derivatives}
\hspace{5mm}
The low energy effective theory on the D-brane is expected to be some kind of 
noncommutative gauge theory.  To construct such a theory we need to define 
proper derivatives. Because the {\boldmath $\star$} product (\ref{newstar}) 
depends on the coordinates explicitly, this will not satify Leibnitz rule.
\begin{equation}
\Delta \equiv \frac{\partial}{\partial x^{\mu}} (f_1 \bm{\star} f_2) -
\left(\frac{\partial}{\partial x^{\mu}} f_1 \right) \bm{\star} f_2 -
f_1 \bm{\star} \left(\frac{\partial}{\partial x^{\mu}}f_2 \right) \neq 0
\label{vLeibnitz}
\end{equation}

To circumvent this problem we must modify the derivatives.  We define
\begin{eqnarray}
\frac{\nabla}{\nabla x^{\mu}} f(x,y) & \equiv&
\frac{\partial}{\partial x^{\mu}} f(x,y) + 
\left\{ {a_{\mu}}^{\nu},\frac{\partial f}{\partial x^{\nu}} \right\}_* +
\left\{ {b_{\mu}}^{\nu},\frac{\partial f}{\partial y^{\nu}} \right\}_*,
\label{derx}
\\
\frac{\nabla}{\nabla y^{\mu}} f(x,y) & \equiv& \frac{\partial}{\partial 
y^{\mu}} f(x,y) + \left\{ {\overline{a}_{\mu}}^{\nu},\frac{\partial f}
{\partial y^{\nu}} \right\}_* + \left\{ {\overline{b}_{\mu}}^{\nu},
\frac{\partial f}{\partial x^{\nu}} \right\}_*.
\label{dery}
\end{eqnarray}
Here ${a_{\mu}}^{\nu}$, ${b_{\mu}}^{\nu}$, ${\overline{a}_{\mu}}^{\nu}$ and 
${\overline{b}_{\mu}}^{\nu}$ are linear functions of $x$, $y$ and are 
${\cal O}(H^1)$. $\{ \ \cdot \  \}_*$ is the anti-commutator with respect 
to the product (\ref{oldstar}). We note that because $a$, $b$, $\overline{a}$, 
$\overline{b}$ are linear functions, $\{a, \partial f/\partial x \}_*$ can 
be replaced by $2a \partial f/\partial x$ in (\ref{derx}), (\ref{dery}).

These functions must be determined in such a way that the new derivatives 
(\ref{derx}), (\ref{dery}) satisfy Leibnitz rule. 
By (\ref{derx}) we obtain 
\begin{eqnarray}
\frac{\nabla}{\nabla x^{\mu}}(f_1 \bm{\star} f_2) - 
\frac{\nabla f_1 }{\nabla x^{\mu}} \bm{\star} f_2 -
f_1 \bm{\star} \frac{\nabla f_2 }{\nabla x^{\mu}}  
&= &  \left[ {a_{\mu}}^{\nu}, f_1\right]_* * \frac{\partial f_2}{\partial
x^{\nu}} - \frac{\partial f_1}{\partial x^{\nu}}* \left[ {a_{\mu}}^{\nu}, 
f_2\right]_* \nonumber \\  &&
+ \left[ {b_{\mu}}^{\nu}, f_1\right]_* * \frac{\partial f_2}{\partial y^{\nu}}
- \frac{\partial f_1}{\partial y^{\nu}}* \left[ {b_{\mu}}^{\nu}, f_2\right]_*
+\Delta
\label{Leibnitz}
\end{eqnarray}
Here $[ \ \cdot \  ]_*$ is the commutator with respect to the product 
(\ref{oldstar}). $\Delta$  can be computed from the 
defintions (\ref{newstar}), (\ref{vLeibnitz}). By requiring 
(\ref{Leibnitz}) to vanish we obtain the equations.
\begin{eqnarray}
&&\theta^{\lambda \rho} \frac{\partial {a_{\mu}}^{\nu}}{\partial x^{\rho}}-
\theta^{\nu \rho} \frac{\partial {a_{\mu}}^{\lambda}}{\partial x^{\rho}} =
-\frac{1}{9}G^{\lambda \alpha}G^{\nu \beta} H_{\alpha \beta \mu} 
 -\frac{1}{18} (\theta^{\lambda \alpha}G^{\nu \beta} + 
G^{\lambda \alpha}\theta^{\nu \beta})H_{\alpha \beta \rho}
{(G^{-1}\theta^{-1})^{\rho}}_{\mu}, \nonumber \\
\end{eqnarray}
\begin{eqnarray}
\theta^{\lambda \rho} \frac{\partial {a_{\mu}}^{\nu}}{\partial y^{\rho}}+
\theta^{\nu \rho} \frac{\partial {b_{\mu}}^{\lambda}}{\partial x^{\rho}}
&=&-\frac{1}{36}(G^{\nu \alpha }G^{\lambda \beta} 
-3\theta^{\nu \alpha }\theta^{\lambda \beta})  H_{\alpha \beta \mu} 
\nonumber \\ &&  \qquad 
 +\frac{1}{36} (\theta^{\nu \alpha}G^{\lambda \beta} + 
G^{\nu \alpha}\theta^{\lambda \beta})H_{\alpha \beta \rho}
{(G^{-1}\theta^{-1})^{\rho}}_{\mu}, 
\end{eqnarray}
\begin{eqnarray}
\theta^{\lambda \rho} \frac{\partial {b_{\mu}}^{\nu}}{\partial y^{\rho}}-
\theta^{\nu \rho} \frac{\partial {b_{\mu}}^{\lambda}}{\partial y^{\rho}}
&=& \frac{1}{18}(G^{\nu \alpha }G^{\lambda \beta} 
-3\theta^{\nu \alpha }\theta^{\lambda \beta})  H_{\alpha \beta \mu} \nonumber 
\\ && \qquad -\frac{1}{18} (\theta^{\nu \alpha}G^{\lambda \beta} + 
G^{\nu \alpha}\theta^{\lambda \beta})H_{\alpha \beta \rho}
{(G^{-1}\theta^{-1})^{\rho}}_{\mu}, 
\end{eqnarray}

General solution to these equations is given by
\begin{eqnarray}
{a_{\mu}}^{\nu}(x,y) &=& -\frac{1}{12}x^{\sigma}(\theta^{-1})_{\sigma \lambda}
(G^{\lambda \alpha}G^{\nu \beta}- \theta^{\lambda \alpha}\theta^{\nu \beta})
H_{\alpha \beta \mu} \nonumber \\
&&+\frac{1}{72}(2x^{\sigma}+y^{\sigma})(\theta^{-1})_{\sigma \lambda}
(G^{\lambda \alpha}G^{\nu \beta}- 3\theta^{\lambda \alpha}\theta^{\nu \beta})
H_{\alpha \beta \mu} \nonumber   \\
&&-\frac{1}{72}(2x^{\sigma}+y^{\sigma})(\theta^{-1})_{\sigma \lambda}
(\theta^{\lambda \alpha}G^{\nu \beta}+G^{\lambda \alpha}\theta^{\nu \beta})
H_{\alpha \beta \rho} {(G^{-1}\theta^{-1})^{\rho}}_{\mu} \nonumber \\
&& -x^{\sigma}(\theta^{-1})_{\sigma \lambda}\ {S^{\lambda \nu}}_{\mu}
-y^{\sigma}(\theta^{-1})_{\sigma \lambda}\ {U^{\lambda \nu}}_{\mu},
\label{a}
\end{eqnarray}
\begin{eqnarray}
{b_{\mu}}^{\nu}(x,y) &=& 
-\frac{1}{72}(x^{\sigma}+2y^{\sigma})(\theta^{-1})_{\sigma \lambda}
(G^{\lambda \alpha}G^{\nu \beta}- 3\theta^{\lambda \alpha}\theta^{\nu \beta})
H_{\alpha \beta \mu} \nonumber   \\
&&+\frac{1}{72}(x^{\sigma}+2y^{\sigma})(\theta^{-1})_{\sigma \lambda}
(\theta^{\lambda \alpha}G^{\nu \beta}+G^{\lambda \alpha}\theta^{\nu \beta})
H_{\alpha \beta \rho} {(G^{-1}\theta^{-1})^{\rho}}_{\mu} \nonumber \\
&& -y^{\sigma}(\theta^{-1})_{\sigma \lambda}\ {T^{\lambda \nu}}_{\mu}
+x^{\sigma}(\theta^{-1})_{\sigma \lambda}\ {U^{\nu \lambda}}_{\mu}.
\label{b}
\end{eqnarray}
Here ${S^{\lambda \nu}}_{\mu} (= {S^{\nu \lambda}}_{\mu})$, 
${T^{\lambda \nu}}_{\mu} (= {T^{\nu \lambda}}_{\mu})$ and 
${U^{\lambda \nu}}_{\mu}$ are some constants to be discussed later. 
Similarly ${\overline{a}_{\mu}}^{\nu}$, ${\overline{b}_{\mu}}^{\nu}$ can be 
determined. By using the $x \leftrightarrow y$ symmetry we find 
\begin{equation}
{\overline{a}_{\mu}}^{\nu}(x,y) ={a_{\mu}}^{\nu}(y,x), \qquad  
{\overline{b}_{\mu}}^{\nu}(x,y) ={b_{\mu}}^{\nu}(y,x).
\end{equation}

The field strength of the gauge fields will be defined as the commutator of 
the gauge covariant derivatives.  For this procedure
to be consistent the derivatives (\ref{derx}), (\ref{dery}) must commute.
Otherwise the commutator becomes a differential operator.
For instance from (\ref{derx}) we obtain
\begin{equation}
\left[ \frac{\nabla}{\nabla x^{\mu}}, \ \frac{\nabla}{\nabla x^{\nu}} \right] 
f = 2\left( \frac{\partial {a_{\nu}}^{\lambda}}{\partial x^{\mu}}-
\frac{\partial {a_{\mu}}^{\lambda}}{\partial x^{\nu}} \right) \ 
\frac{\partial f}{\partial x^{\lambda}} + 2\left( \frac{\partial 
{b_{\nu}}^{\lambda}}{\partial x^{\mu}}-\frac{\partial {b_{\mu}}^{\lambda}}
{\partial x^{\nu}} \right) \ \frac{\partial f}{\partial y^{\lambda}} 
\end{equation}
and the two coefficients of the derivatives of $f$ must vanish. We obtain 4
more conditions from the other commutators.  It turns out the following three 
out of the six conditions are independent.
\begin{equation}
\frac{\partial {a_{\nu}}^{\lambda}}{\partial x^{\mu}} =
\frac{\partial {a_{\mu}}^{\lambda}}{\partial x^{\nu}}, \qquad
\frac{\partial {b_{\nu}}^{\lambda}}{\partial x^{\mu}} =
\frac{\partial {b_{\mu}}^{\lambda}}{\partial x^{\nu}}, \qquad
\frac{\partial {\overline{b}_{\nu}}^{\lambda}}{\partial x^{\mu}} =
\frac{\partial {a_{\mu}}^{\lambda}}{\partial y^{\nu}} 
\label{integrability}
\end{equation}
These equations impose the following conditions on $S$, $T$, $U$.
\begin{eqnarray}
 {S^{\mu \lambda}}_{\rho} \ \theta^{\rho \nu}- 
{S^{\nu \lambda}}_{\rho} \ \theta^{\rho \mu} 
&=& \frac{1}{18} G^{\mu \alpha} G^{\nu \beta} \theta^{\lambda \gamma} \ 
H_{\alpha \beta \gamma} 
 +\frac{1}{36} \left(\theta^{\nu \alpha}G^{\mu \beta}-
\theta^{\mu \alpha}G^{\nu \beta} \right) G^{\lambda \gamma} \ 
H_{\alpha \beta \gamma}, \nonumber \\
&&
\label{1}
\end{eqnarray}
\begin{eqnarray}
(\theta^{-1})_{\mu \rho} \ {U^{\lambda \rho}}_{\nu} - 
(\theta^{-1})_{\nu \rho} \ {U^{\lambda \rho}}_{\mu} 
&=& 
 \frac{1}{36}{\left(G^{-1}\theta^{-1} \right)^{\sigma}}_{\nu} \
 G^{\lambda \beta} H_{\mu \sigma \beta}
-\frac{1}{36}{\left(G^{-1}\theta^{-1} \right)^{\sigma}}_{\mu} \
 G^{\lambda \beta} H_{\nu \sigma \beta}  \nonumber \\
&&- \frac{1}{36} {\left(G^{-1}\theta^{-1} \right)^{\alpha}}_{\mu}
{\left(G^{-1}\theta^{-1} \right)^{\beta}}_{\nu} \theta^{\lambda \sigma}
H_{\alpha \beta \sigma} + \frac{1}{12} \theta^{\lambda \beta} 
H_{\mu \nu \beta}, \nonumber \\
\label{2}
\end{eqnarray}
\begin{eqnarray}
{U^{\nu \sigma}}_{\lambda} &=& \theta^{\nu \alpha} \ 
(\theta^{-1})_{\lambda \beta} {T^{\beta \sigma}}_{\alpha} -\frac{1}{72}\left(
G^{\nu \alpha}G^{\sigma \beta}-3\theta^{\nu \alpha} \theta^{\sigma \beta} 
\right) \ H_{\lambda \alpha \beta} \nonumber \\
&& \qquad \qquad \qquad \qquad \qquad \qquad + \frac{1}{72} \left( 
\theta^{\nu \alpha}G^{\sigma \beta} +  G^{\nu \alpha}\theta^{\sigma \beta} 
\right) \ {\left(G^{-1}\theta^{-1} \right)^{\rho}}_{\lambda} \ H_{\alpha 
\beta \rho} \nonumber \\
\label{5}
\end{eqnarray}
If we recall the symmetry ${T^{\lambda \nu}}_{\mu} = {T^{\nu \lambda}}_{\mu}$,
(\ref{2}) comes out as a result of (\ref{5}) and hence is not an independent
condition.

The solution to (\ref{1}) is given by 
\begin{eqnarray}
{S^{\mu \nu}}_{\lambda}&=& c \left(\theta^{\mu \alpha}G^{\nu \beta}-
G^{\mu \alpha} \theta^{\nu \beta} \right) \ H_{\alpha \beta \rho} \ 
{(\theta G)^{\rho}}_{\lambda} +c \left\{ (\theta G \theta)^{\mu \alpha}
G^{\nu \beta}+(\theta G \theta)^{\nu \alpha}G^{\mu \beta} \right\} \ 
H_{\alpha \beta \lambda} \nonumber \\
&& +c \left\{ (\theta G \theta)^{\mu \alpha}\theta^{\nu \beta}+
(\theta G \theta)^{\nu \alpha}\theta^{\mu \beta} \right\} \ 
H_{\alpha \beta \rho} \ {(G^{-1}\theta^{-1})^{\rho}}_{\lambda} \nonumber \\
&&+\frac{1}{36}  (\theta^{\mu \alpha}
G^{\nu \beta}-G^{\mu \alpha}\theta^{\nu \beta}) \ H_{\alpha \beta \rho} \ 
{(G^{-1}\theta^{-1})^{\rho}}_{\lambda}, 
\label{STU}
\end{eqnarray}
$c$ is a constant.  ${T^{\mu \nu}}_{\lambda}$ and ${U^{\mu \nu}}_{\lambda}$ 
are arbitrary as long as (\ref{5}) is satisfied. Unfortunately, further 
conditions which may determine these values are not yet available to the 
authors. 

The fact that the derivatives (\ref{derx}), (\ref{dery}) commute suggests  
the existence of a coordinate transformation $x,y \rightarrow x', y'$
by which the derivatives (\ref{derx}), (\ref{dery}) are reexpressed as ordinary
derivatives.
\begin{equation}
\frac{\partial}{\partial x^{\prime \mu}} = 
\frac{\nabla}{\nabla x^{\mu}} = \frac{\partial}{\partial x^{\mu}}
+ 2{a_{\mu}}^{\nu} \ \frac{\partial}{\partial x^{\nu}}
+ 2{b_{\mu}}^{\nu} \ \frac{\partial}{\partial y^{\nu}}
\label{derivative}
\end{equation} 
(And a similar equation for $\nabla / \nabla y$.)  Here we used the property, 
$\{a, h \}_* = 2a h$ which holds for a linear function $a$. In this 
transformation $x$, $y$ are treated as commuting variables. 

Indeed eq (\ref{integrability}) guarantees the existence of such a 
transformation and it is given by
\begin{eqnarray}
x^{\prime \nu} &=& x^{\nu} + \frac{1}{18} x^{\sigma}x^{\mu} 
(\theta^{-1})_{\sigma \lambda} G^{\lambda \alpha}G^{\nu \beta} \ 
H_{\alpha \beta \mu} \nonumber \\
&& + \frac{1}{36} x^{\sigma}x^{\mu} (\theta^{-1})_{\sigma \lambda}
 (\theta^{\lambda \alpha}G^{\nu \beta} +G^{\lambda \alpha}
\theta^{\nu \beta}) {(G^{-1}\theta^{-1})^{\rho}}_{\mu}
\ H_{\alpha \beta \rho} \nonumber \\
&& - \frac{1}{36} y^{\sigma}x^{\mu} (\theta^{-1})_{\sigma \lambda}
 (G^{\lambda \alpha}G^{\nu \beta} -3\theta^{\lambda \alpha}
\theta^{\nu \beta}) \ H_{\alpha \beta \mu} \nonumber \\
&& + \frac{1}{36} y^{\sigma}x^{\mu} (\theta^{-1})_{\sigma \lambda}
 (\theta^{\lambda \alpha}G^{\nu \beta} +G^{\lambda \alpha}
\theta^{\nu \beta}) {(G^{-1}\theta^{-1})^{\rho}}_{\mu}
\ H_{\alpha \beta \rho} \nonumber \\
&& + \frac{1}{72} y^{\sigma}y^{\mu} (\theta^{-1})_{\sigma \lambda}
 (G^{\lambda \alpha}G^{\nu \beta} -3\theta^{\lambda \alpha}
\theta^{\nu \beta}) \ H_{\alpha \beta \mu} \nonumber \\
&& - \frac{1}{72} y^{\sigma}y^{\mu} (\theta^{-1})_{\sigma \lambda}
 (\theta^{\lambda \alpha}G^{\nu \beta} +G^{\lambda \alpha}
\theta^{\nu \beta}) {(G^{-1}\theta^{-1})^{\rho}}_{\mu}
\ H_{\alpha \beta \rho} \nonumber \\
&&+x^{\sigma}x^{\mu}(\theta^{-1})_{\sigma \lambda}{S^{\lambda \nu}}_{\mu}
+2x^{\mu}y^{\sigma}(\theta^{-1})_{\sigma \lambda}{U^{\lambda \nu}}_{\mu}
-y^{\sigma}y^{\mu}(\theta^{-1})_{\sigma \lambda}{U^{\nu \lambda}}_{\mu}
\label{transfx}
\end{eqnarray}
and a similar equation with the replacement
$x \leftrightarrow y, \ \ x' \leftrightarrow y'$.

By using (\ref{CR1})-(\ref{CR3}) it is possible to show that in the primed 
coordinate system $x'$ and $y'$ commute with each other.
\begin{equation}
\ [x^{\prime \mu}, x^{\prime \nu} ]=i \theta^{\mu \nu}, \qquad 
[y^{\prime \mu}, y^{\prime \nu} ]=-i \theta^{\mu \nu}, \qquad 
[x^{\prime \mu}, y^{\prime \nu} ]=0
\end{equation}
Here the quadratic terms in (\ref{transfx}) are assumed to be symmetrized. 
In other words the effect of the field strength $H$ can be compensated by 
a `coordinate transformation'.  This fact suggests that the metric tensor in 
the `$(x',y')$ space' may be also flat and given by $G_{\mu \nu}$. We should 
note that this is not the ordinary space but a dimensionally-doubled space 
combining the coordinates of the two ends of the string and that the 
transformation (\ref{transfx}) mixes up $x$ and $y$ as a whole. It is 
tempting to assume that the existence of such a transformation persists to 
all orders of $H$. We do not have a proof at present.

Let us construct a trace operation $Tr f(x,y)$.  This must satisfy the
cyclic property.
\begin{equation}
Tr \ f_1 \bm{\star} f_2 = Tr \ f_2 \bm{\star} f_1
\label{cyclic}
\end{equation}
We assume the form 
\begin{equation}
Tr \  f \equiv \int d^Dx d^Dy \  G \  \Omega(x,y)  \ f(x,y)
\label{trace}
\end{equation}
and determine the function $\Omega(x,y)$ by the condition
\begin{equation}
\int d^Dx d^Dy \ G\  \Omega  \ (f_1 \bm{\star} f_2) =
\int d^Dx d^Dy \ G\  \Omega \  (f_2 \bm{\star} f_1).
\label{cyclicproperty}
\end{equation}
For $H=0$ we must also have $\Omega =1$.
Note that on the righthand side of (\ref{trace}) the product is the 
ordinary one. $G$ is $\det G_{\mu \nu}$. It is not hard to show that this 
function is given by
\begin{equation}
\Omega(x,y) = 1 - \frac{1}{6}(x^{\sigma}+y^{\sigma}) \ \theta^{\mu \lambda} \ 
H_{\mu \lambda \sigma} -\frac{1}{6}(x^{\sigma}+y^{\sigma}) \ 
(G^{-1}\theta^{-1} G^{-1})^{\mu \lambda} \ H_{\mu \lambda \sigma}.
\end{equation}

It may be natural to expect that this extra factor $\Omega$ in the trace
has an origin in the Jacobian from $(x, y)$ to $(x',y')$. We indeed obtain 
\begin{eqnarray}
\det \frac{(\partial x', \partial y')}{(\partial x, \partial y)} &=& 
1 - \frac{1}{12}(x^{\sigma}+y^{\sigma}) \ 
\theta^{\mu \lambda} \ H_{\mu \lambda \sigma} 
-\frac{1}{12}(x^{\sigma}+y^{\sigma}) \ 
(G^{-1}\theta^{-1} G^{-1})^{\mu \lambda} \ H_{\mu \lambda \sigma} \nonumber \\
&& \qquad \qquad +2(x^{\sigma}+y^{\sigma})(\theta^{-1})_{\sigma 
\lambda}({S^{\lambda \mu}}_{\mu}+{U^{\lambda \mu}}_{\mu}) 
\end{eqnarray}
and we can show from (\ref{1})-(\ref{5}) that this coincides with $\Omega$.

\section{Noncommutative gauge theory for $H \neq 0$}
\hspace{5mm}
We are now ready to define gauge transformations and write down the gauge 
theory action. This action will be an integral over $x^{\mu}$ and $y^{\mu}$
and defines a bi-local theory. We will construct this bi-local theory as an 
intermediate step for obtaining a `local gauge theory' in the next section.  

Let us define the gauge covariant derivatives  by 
\begin{eqnarray}
D_{\mu} &=& \partial / \partial x^{\prime \mu} + i A_{\mu}' =
\nabla / \nabla x^{\mu} + i A_{\mu}', \nonumber \\
\overline{D}_{\mu} &=& \partial / \partial y^{\prime \mu} - i 
\overline{A}_{\mu}'=
\nabla / \nabla y^{\mu} - i \overline{A}_{\mu}'.
\label{covder}
\end{eqnarray}
Here $A_{ \mu}'$, $\overline{A}_{\mu}'$ are the gauge fields in the $(x',y')$ 
frame and transformed to those in the $(x,y)$ frame like the derivatives
 (\ref{derivative}),
\begin{eqnarray}
A_{ \mu}' & =&  A_{ \mu} + 2{a_{\mu}}^{\nu} A_{\nu} - 2{b_{\mu}}^{\nu} 
\overline{A}_{\nu}, \nonumber \\
\overline{A}_{ \mu}' & =&  \overline{A}_{ \mu} + 2\overline{a}_{\mu}{}^{\nu} 
\overline{A}_{\nu} -
 2\overline{b}_{\mu}{}^{\nu} A_{\nu}.
\label{A} 
\end{eqnarray}
We will use a `bar' to denote functions associated with the coordinate $y$. 
In (\ref{covder}), (\ref{A}) the signs in front of $A_{ \mu}'$ and 
$\overline{A}_{\mu}'$ are opposite, because
the two ends of the open string have opposite charges. 
The gauge transformation is defined by 
\begin{eqnarray}
D_{\mu} & \rightarrow & U \bm{\star} D_{\mu} \bm{\star} U^{-1}, \qquad 
\overline{D}_{\mu} \rightarrow  U \bm{\star} \overline{D}_{\mu} \bm{\star} 
U^{-1}, \qquad \Phi  \rightarrow  U \bm{\star} \Phi \bm{\star} U^{-1},
\label{gaugetr}
\end{eqnarray}
where $U = \exp_{\bm{\star}} (i \Lambda)$ is a gauge function.
$\exp_{\bm{\star}}$ is an exponential defined in terms of {\boldmath $\star$}. 
$\Phi$ is a scalar field in the adjoint representation.
For an infinitesimal $\Lambda$ these read
\begin{eqnarray}
\delta A_{\mu}' & =& i [ \Lambda,  A_{\mu}']_{\star} 
- \frac{\nabla \Lambda} {\nabla x^{\mu}}, \qquad 
\delta \overline{A}_{\mu}' = i [ \Lambda,  \overline{A}_{\mu}']_{\star} + 
\frac{\nabla \Lambda}{\nabla y^{\mu}}, \qquad 
\delta \Phi = i [ \Lambda,  \Phi]_{\star}. 
\label{infgauge}
\end{eqnarray}
Here $[ \ \cdot \  ]_{\star}$ is the commutator with respect to the product
(\ref{newstar}). An infinitesimal transformation of $A_{\mu}$, 
$\overline{A}_{\mu}$ is consistently obtained. 
\begin{eqnarray}
\delta A_{\mu}& =& i [ \Lambda,  A_{\mu}]_{\star} - \frac{\partial \Lambda}
{\partial x^{\mu}} + i \{ [\Lambda, \ {a_{\mu}}^{\lambda} ]_{\star}, \  
A_{\lambda} \}_{\star}  - i \{ [\Lambda, \ {b_{\mu}}^{\lambda} ]_{\star}, \  
\overline{A}_{\lambda} \}_{\star}, 
\end{eqnarray}
\begin{eqnarray}
\delta \overline{A}_{\mu}& =& i [ \Lambda,  \overline{A}_{\mu}]_{\star} +  
\frac{\partial \Lambda} {\partial y^{\mu}} + i \{ [\Lambda, \ 
\overline{a}_{\mu}{}^{\lambda} ]_{\star}, \  
\overline{A}_{\lambda} \}_{\star}  
- i \{ [\Lambda, \ \overline{b}_{\mu}{}^{\lambda} ]_{\star}, \  A_{\lambda} 
\}_{\star}
\end{eqnarray}

Now by using the standard prescription we can write down the action integral 
for the bi-local noncommutative gauge theory
\begin{eqnarray}
S_{\rm{bi-local}} &=&  \frac{1}{4g^2_{\rm{YM}}V}\int d^Dx d^Dy \ G\ 
\Omega(x,y) \ \nonumber \\
&& \qquad \qquad \cdot \left\{ \ G^{\mu \nu} G^{\lambda \rho} \  
[D_{\mu}, \  D_{\lambda}
]_{\star} \ \bm{\star} \  [D_{\nu}, \ D_{\rho} ]_{\star} 
- 2 G^{\mu \nu} \ [D_{\mu}, \  X^m ]_{\star} 
\bm{\star} [D_{\nu}, \ X^m ]_{\star} \right. \nonumber \\
&& \qquad \qquad \qquad + \ [X^m, \ X^n]_{\star}^2  \
+ \ G^{\mu \nu} G^{\lambda \rho} \  [\overline{D}_{\mu}, 
\ \overline{D}_{\lambda} ]_{\star} \ \bm{\star} \  [\overline{D}_{\nu}, 
\ \overline{D}_{\rho} ]_{\star} \nonumber \\ 
&& \qquad \qquad \qquad \left.  - 2 \ G^{\mu \nu} \ [\overline{D}_{\mu}, \  
\overline{X}^m ]_{\star} \bm{\star} [\overline{D}_{\nu}, \ 
\overline{X}^m ]_{\star}  + \ [\overline{X}^m, \ 
\overline{X}^n]_{\star}^2 \  \right\} \nonumber \\
& \equiv & \int d^Dx d^Dy \ G V^{-1} \ \Omega \ {\cal L}(x,y),
\label{gauge}
\end{eqnarray} 
where $V$ is the volume $\int d^Dx \ \sqrt{G}$ and $g_{\rm{YM}}$ the 
Yang-Mills coupling. We explicitly write only the integration measure 
relevant to our discussions. $X^m$ and $\overline{X}^m$ are the scalar fields 
describing the fluctuations of the D-brane.  The scalar fields are doubled, 
because there are two such independent fluctuations at both ends of the open 
string. This action is invariant under the gauge transformation 
(\ref{gaugetr}) due to the cyclic property (\ref{cyclicproperty}) of the trace 
(\ref{trace}).  Because $G^{\mu \nu}$ is constant, this does not break the 
gauge symmetry. There is no constraint on the gauge function $\Lambda(x,y)$.

By substituting (\ref{derx}), (\ref{dery}), (\ref{covder}), (\ref{A}) into 
(\ref{gauge}) we will find a non-constant metric in the action. For example 
we obtain for the commutator of the covariant derivatives
\begin{eqnarray}
\ [ \ D_{\mu}, \ D_{\nu} \ ]_{\star} &=& i \ \{ {\cal F}_{\mu \nu}^{(x,x)}
+ 2{a_{\mu}}^{\lambda} \ {\cal F}_{\lambda \nu}^{(x,x)}
+ 2{a_{\nu}}^{\lambda} \ {\cal F}_{\mu \lambda}^{(x,x)}
+ 2{b_{\mu}}^{\lambda} \ {\cal F}_{\lambda \nu}^{(y,x)} \nonumber \\
&& \qquad + 2{b_{\nu}}^{\lambda} \ {\cal F}_{\mu \lambda}^{(x,y)} 
+ {\cal R}_{\mu \nu} \},
\end{eqnarray}
where ${\cal F}$'s are the field strength in terms of the ordinary derivatives.
\begin{eqnarray}
{\cal F}_{\mu \nu}^{(x,x)} &=& \frac{\partial}{\partial x^{\mu}}A_{\nu}
-\frac{\partial}{\partial x^{\nu}}A_{\mu}
+i \ [ \ A_{\mu}, \ A_{\nu} \ ]_{\star}, \nonumber \\
{\cal F}_{\mu \nu}^{(y,x)} &=& -\frac{\partial}{\partial y^{\mu}}A_{\nu}
-\frac{\partial}{\partial x^{\nu}}\overline{A}_{\mu}
+i \ [ \ \overline{A}_{\mu}, \ A_{\nu} \ ]_{\star}
= {\cal F}_{\nu \mu}^{(x,y)}, \nonumber \\
{\cal F}_{\mu \nu}^{(y,y)} &=& \frac{\partial}{\partial y^{\mu}}
\overline{A}_{\nu}-\frac{\partial}{\partial y^{\nu}}\overline{A}_{\mu}
- i \ [ \ \overline{A}_{\mu}, \ \overline{A}_{\nu} \ ]_{\star}
\end{eqnarray}
${\cal R}_{\mu \nu}$ is a sum of terms like 
\begin{equation}
-i\theta^{\rho \sigma} \frac{\partial {a_{\mu}}^{\lambda}}{\partial x^\rho} 
\left\{A_{\lambda}, \ \frac{\partial A_{\nu}}{\partial x^\sigma} 
\right\}_{\star}.
\end{equation}
Here we omit its explicit form. By defining a new curved metric
\begin{eqnarray}
\tilde{G}^{(x,x)\mu \nu}(x,y) &=& G^{\mu \nu} + 2G^{\mu \lambda} \ 
{a_{\lambda}}^{\nu}  + 2G^{\nu \lambda} \ {a_{\lambda}}^{\mu}, \nonumber \\
\tilde{G}^{(y,y)\mu \nu}(x,y) &=& G^{\mu \nu} + 2G^{\mu \lambda} \ 
{\overline{a}_{\lambda}}^{\nu}
 + 2G^{\nu \lambda} \ {\overline{a}_{\lambda}}^{\mu}, \nonumber \\
\tilde{G}^{(x,y)\mu \nu}(x,y) &=& 4G^{\mu \lambda} \ {b_{\lambda}}^{\nu},
\qquad  \tilde{G}^{(y,x)\mu \nu}(x,y) = 4G^{\mu \lambda} \ 
{\overline{b}_{\lambda}}^{\nu},
\label{tildeG}
\end{eqnarray}
we obtain the part of the action for the gauge field.
\begin{eqnarray}
S_{\rm{gauge} \ \rm{field}} &=&  \frac{-1}{4g^2_{\rm{YM}}V}\int d^Dx d^Dy \  
\sqrt{\tilde{G}(x,y)} \ \nonumber \\
&& \cdot \{ \ \tilde{G}^{(x,x)\mu \nu} \cdot \tilde{G}^{(x,x)
\lambda \rho} \cdot {\cal F}_{\mu\lambda}^{(x,x)} \  \bm{\star} \  
{\cal F}_{\nu\rho}^{(x,x)} + \tilde{G}^{(y,y)\mu \nu} \cdot \tilde{G}^{(y,y)
\lambda \rho} \cdot {\cal F}_{\mu\lambda}^{(y,y)} \ \bm{\star} \  
{\cal F}_{\nu\rho}^{(y,y)}  \nonumber \\
&& + 2\tilde{G}^{(x,x)\mu \nu} \cdot \tilde{G}^{(x,y)
\lambda \rho} \cdot  {\cal F}_{\mu\lambda}^{(x,x)} \  \bm{\star} \  
{\cal F}_{\nu\rho}^{(x,y)} + 2 
\tilde{G}^{(y,y)\mu \nu} \cdot \tilde{G}^{(y,x)\lambda \rho} \cdot 
{\cal F}_{\mu\lambda}^{(y,y)} \ \bm{\star} \  
{\cal F}_{\nu\rho}^{(y,x)} \nonumber \\
&&+ 2G^{\mu \lambda} \cdot G^{\nu \rho} \cdot {\cal F}_{\mu \nu}^{(x,x)} 
\bm{\star} {\cal R}_{\lambda \rho}+ 2 G^{\mu \lambda} \cdot G^{\nu \rho} 
\cdot {\cal F}_{\mu \nu}^{(y,y)} \bm{\star} \overline{{\cal R}}_{\lambda 
\rho} \ \}
\end{eqnarray}
Here $\tilde{G}(x,y)$ is the determinant of the 2$D$ $\times$ 2$D$ matrix 
($\tilde{G}^{(x,x)}_{\mu \nu}$, $\tilde{G}^{(x,y)}_{\mu \nu}$  / 
$\tilde{G}^{(y,x)}_{\mu \nu}$, $\tilde{G}^{(y,y)}_{\mu \nu}$) and coincides 
with $\{\Omega(x,y)\}^2$. $\overline{{\cal R}}_{\mu \nu}$ is a sum of terms 
similar to those of ${\cal R}_{\mu \nu}$. In the above `$\cdot$' is an 
ordinary product. 
In this way the noncommutative gauge theory in the curved space is obtained.  
Although the metric $\tilde{G}^{(\cdot \cdot) \mu \nu}$ depends on the 
coordinates, the action is gauge invariant. Similarly, an action for the 
scalars $X$, $\overline{X}$ can be obtained.

In the special case $H =0$, $x$ and $y$ commute and the product of functions 
of $x$ only is still a function of $x$ only. Thus it is natural to set 
$A_{\mu}= A_{\mu}(x)$, $\overline{A}_{\mu}=\overline{A}_{\mu}(y)$, 
$X^m=X^m(x)$, $\overline{X}^m=\overline{X}^m(y)$.  The product {\boldmath 
$\star$} reduces to $*$. Then the action $S_{\rm{bi-local}}$ 
becomes the sum of two decoupled noncommutative gauge theories with 
noncommutative parameters $\theta$ and $-\theta$, respectively. 

In $H=0$ noncommutative gauge theory, one obtains unitary and causal S-matrix
as long as the noncommutative parameter has only space components.\cite{causal}
If we regard (\ref{gauge}) as the action integral in 2$D$-dimensional space, 
we may be able to expect that this theory is also causal, even if there are 
terms containing the field strength $H$. If we regard $x$ and $y$ as two 
distinct points in the same $D$-dimensional space, however, this theory will 
be non-local and we cannot expect to obtain causal S-matrix.  We must reduce 
$S_{\rm{bi-local}}$ to a single $x$-integral and make the theory `local'.

\section{The $y^{\mu} \rightarrow x^{\mu}$ limit and the reduced effective 
action}
\hspace{5mm}
We need to find a prescription to connect this action to the amplitudes of  
the string theory in the low energy limit. Let us recall that the energy of 
the open string is proportional to the length of the string divided by 
$\alpha'$ and given by $M \simeq \frac{1}{\alpha'} \sqrt{G_{\mu \nu} \ p^{\mu} 
\ p^{\nu}} \simeq \frac{1}{\alpha'} \sqrt{G_{\mu \nu} \ 
{(J^{-1})^{\mu}}_{\lambda} \ (y-x)^{\lambda} \ {(J^{-1})^{\nu}}_{\rho} \ 
(y-x)^{\rho} }$. In the low energy limit this goes to a 
finite value. A massive string will be unstable and eventually decay into a 
massless string. To obtain a massless string we must take the $y \rightarrow 
x$ limit by hand. If one sends $y$ to $x$ directly in the product {\boldmath 
$\star$}, (\ref{newstar}), this product becomes non-associative. In this 
sense the separation $y-x$ works as a kind of regularization to realize the 
associative product {\boldmath $\star$}. At the final step we need to reduce 
the $x$, $y$ integral to a single $x$ integral.  The integration measure is 
not just $ d^Dx \ \Omega (x,x)$ but must be modified. 

As a simple example let us first consider the following integral in the 
$(x,y)$ space.
\begin{equation}
\int d^Dx d^Dy \ \Omega(x,y) \ \Phi_1(x,y) \bm{\star} \Phi_2(x,y) \cdots 
\bm{\star} \Phi_n(x,y) 
\end{equation}
$\Phi_i(x,y)$'s are scalar fields.  We propose that the following 
reduced integral in the $x$ space should correspond to this integral
in the $y \rightarrow x$ limit.\footnote{At ${\cal O}(H^1)$ this prescription 
is similar to the one ($p=0$) adopted in \cite{Ho3} due to the relation 
(\ref{pyx}).}
\begin{equation}
\int d^Dx  \ \omega(x)\ [ \ 
\Phi_1(x,y) \bm{\star} \Phi_2(x,y) \cdots \bm{\star} \Phi_n(x,y)  \ ]|_{y=x}
\label{reduced}
\end{equation}
In (\ref{reduced}) we set $y=x$ only after all the algebra associated with 
{\boldmath $\star$} is finished.  Here $\omega(x)$ is a function which is  
determined by the requirement of cyclic property.  Actually if we set 
\begin{equation}
\omega(x) = 1 + \frac{1}{3} x^{\lambda} \  \theta^{\mu \nu} \ 
H_{\lambda \mu \nu},
\label{omega}
\end{equation}
the two point function of the functions of $x$ only, 
\begin{equation}
I_2 [\Phi_1,\Phi_2] \equiv \int d^Dx \ \omega(x) \left. (\Phi_1(x) \bm{\star} 
\Phi_2(x)) \right|_{y=x}
\end{equation}
is given by 
\begin{equation}
I_2[\Phi_1,\Phi_2] = \int d^Dx \Phi_1 (x) \cdot \Phi_2(x) \cdot 
(1 + \frac{1}{3} x^{\lambda} \theta^{\mu \nu} \ H_{\mu \nu \lambda}).
\label{I2}
\end{equation}
This is cyclic invariant.
\begin{equation}
I_2[\Phi_1,\Phi_2] =I_2[\Phi_2,\Phi_1]
\label{cyc2}
\end{equation}
Contrary to our expectation the function $\omega(x)$ is not proportional to 
$\sqrt{\det \hat{G}_{\mu \nu}(x)}=\sqrt{G} \ ( 1 - \frac{1}{3} x^{\lambda} \  
\theta^{\mu \nu} \ H_{\lambda \mu \nu} )$ but to $\sqrt{\det \hat{G}^{\mu 
\nu}(x)}$, where $\hat{G}_{\mu \nu}(x)=(g \ (1-(g^{-1}B(x))^2))_{\mu \nu}$ 
is the open string metric in the presence of $H$.

Furthermore we can check that this correlation function enjoys the following 
interesting property.
\begin{eqnarray}
&& \int d^Dx \ \omega \left. (\Phi_1(x) \bm{\star} \Phi_2(x)) \right|_{y=x} 
= \int d^Dx \ \omega \left. (\Phi_1(x) \bm{\star} \Phi_2(y)) \right|_{y=x} 
\nonumber \\
&=& \int d^Dx \ \omega \left. (\Phi_1(y) \bm{\star} \Phi_2(x)) \right|_{y=x}
= \int d^Dx \ \omega \left. (\Phi_2(y) \bm{\star} \Phi_1(y)) \right|_{y=x}
\label{cycl2}
\end{eqnarray}
This property reveals the string theory origin of the {\boldmath $\star$} 
product. To see this it is helpful to visualize the {\boldmath $\star$} 
product as an operation adding one rung to a ladder. Let us consider a ladder 
and name the two legs as $x$ and $y$, respectively.  The $\bm{\star} $ 
product adds one unit on top of the ladder. $\bm{\star} \ \Phi_i(x)$ adds 
one rung with the $x$-leg marked, while $\bm{\star} \  \Phi_i(y)$ one with the 
$y$-leg marked.  This operation must be performed step by step and is 
noncommutative.  Finally setting $y=x$ and integrating over $x$ join the two 
legs at the top and bottom, separately. After this joining the legs are 
merged into a circle and we expect to have the cyclic property as in the 
string amplitude. This is the case here.  Eq (\ref{cycl2}) is illustrated 
in Fig.1.

\vspace*{1cm } 
\newlength{\haba}
\newlength{\backu}
\setlength{\backu}{\unitlength}
\unitlength 1.3mm
\setlength{\haba}{10\unitlength}
\begin{tabular}{ccccccc}
\begin{picture}(3,0)(0,0)
 \put(1,-1){\makebox(0,0){$\Phi_1(x)$}}
 \put(1,3){\makebox(0,0){$\Phi_2(x)$}}
\end{picture}
\begin{picture}(0,0)(0,0)
\put(1,-5){\makebox(0,0){$x$}}
\put(10,-5){\makebox(0,0){$y$}}
\end{picture}
\begin{minipage}[c]{\haba}
\epsfxsize=\haba \epsfbox {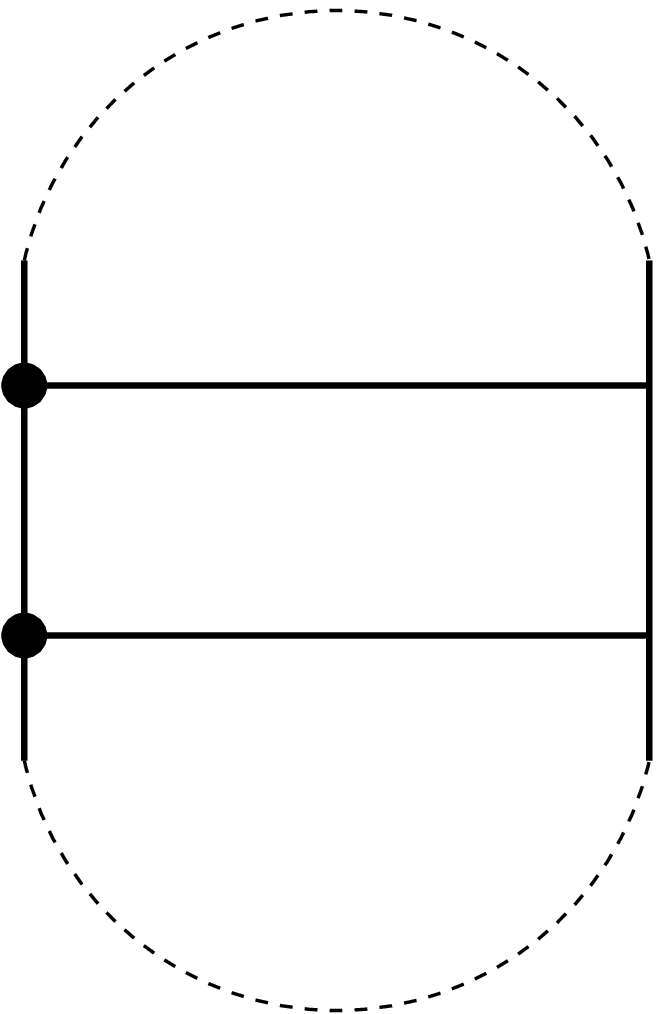}
\end{minipage}
 &=&
\begin{picture}(3,0)(0,0)
 \put(1.5,-1){\makebox(0,0){$\Phi_1(x)$}}
\end{picture}
\begin{picture}(0,0)(0,0)
\put(1,-5){\makebox(0,0){$x$}}
\put(10,-5){\makebox(0,0){$y$}}
\end{picture}
\begin{minipage}[c]{\haba}
\epsfxsize=\haba \epsfbox {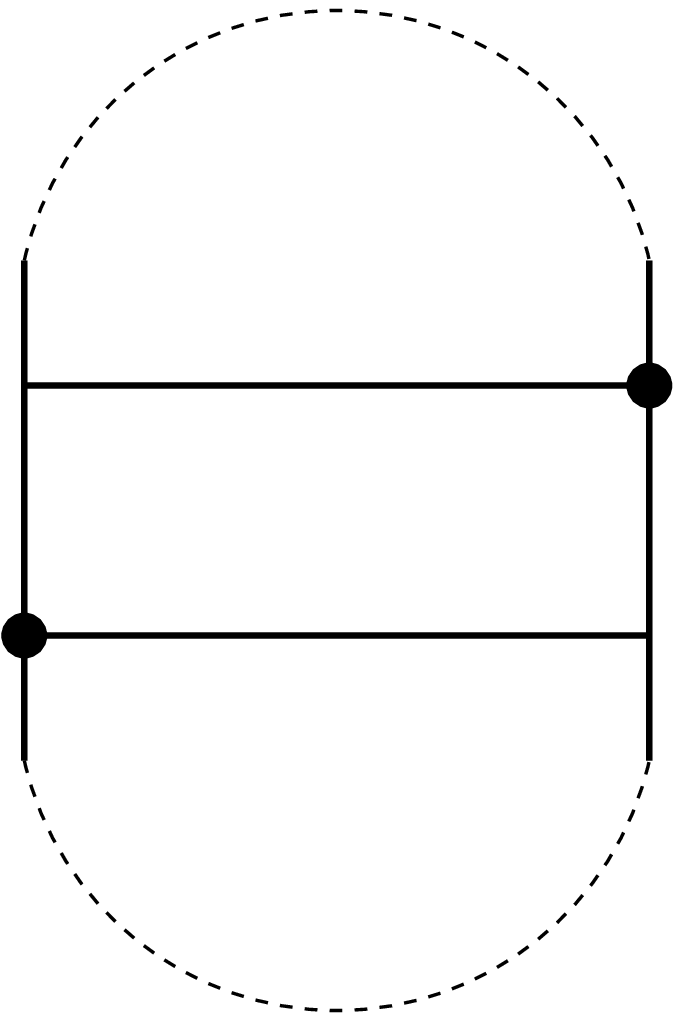}  
\end{minipage}
\begin{picture}(3,0)(0,0)
 \put(2.5,3){\makebox(0,0){$\Phi_2(y)$}}
\end{picture}
 & = &
\begin{picture}(3,0)(0,0)
 \put(1.5,3){\makebox(0,0){$\Phi_1(x)$}}
\end{picture}
\begin{picture}(0,0)(0,0)
\put(1,-5){\makebox(0,0){$x$}}
\put(10,-5){\makebox(0,0){$y$}}
\end{picture}
\begin{minipage}[c]{\haba}
\epsfxsize=\haba \epsfbox {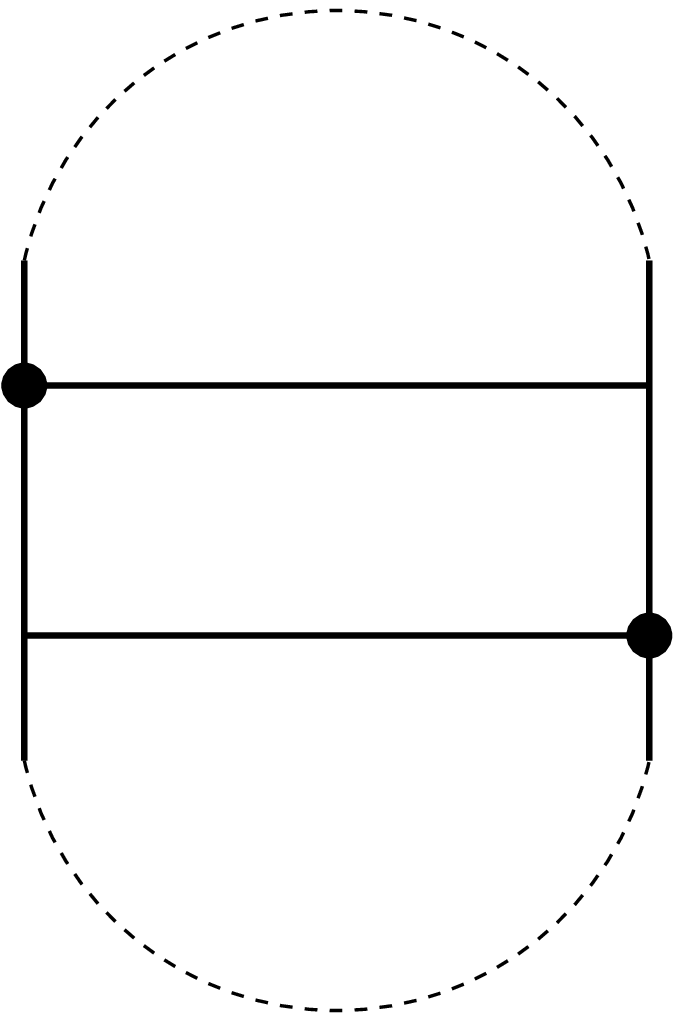}
\end{minipage}
\begin{picture}(3,0)(0,0)
 \put(2.5,-1){\makebox(0,0){$\Phi_2(y)$}}
\end{picture}
 & = &
\begin{picture}(0,0)(0,0)
\put(1,-5){\makebox(0,0){$x$}}
\put(10,-5){\makebox(0,0){$y$}}
\end{picture}
\begin{minipage}[c]{\haba}
 \epsfxsize=\haba \epsfbox {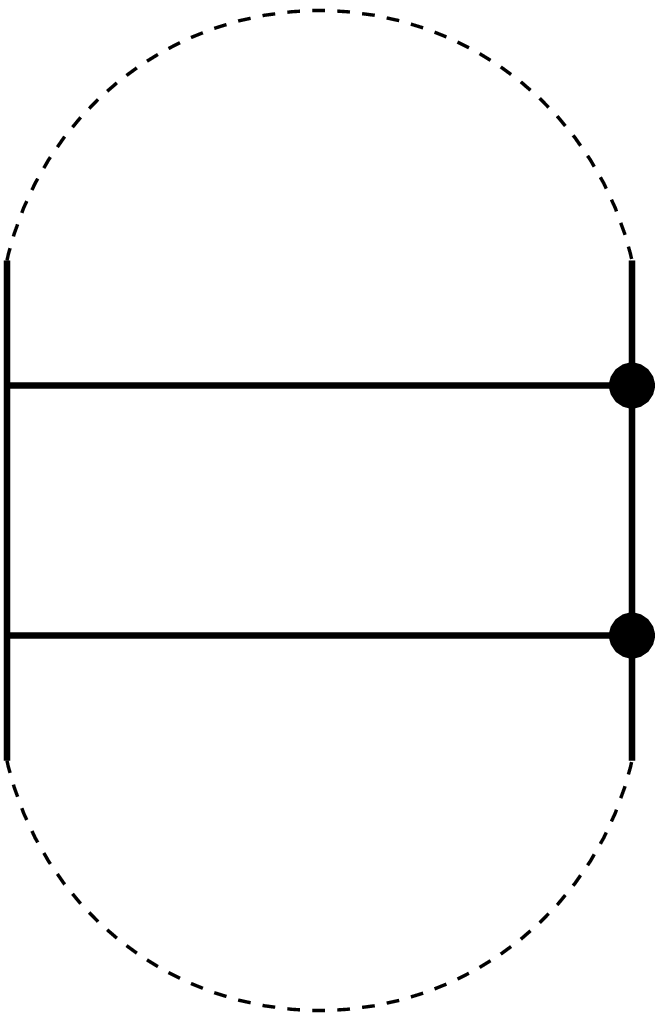}
\end{minipage}
\begin{picture}(3,0)(0,0)
 \put(2.5,3){\makebox(0,0){$\Phi_1(y)$}}
 \put(2.5,-1){\makebox(0,0){$\Phi_2(y)$}}
\end{picture}
\\
\makebox[3\unitlength]{}$\Phi_1(x)\star\Phi_2(x)$
&
&
$\Phi_1(x)\star\Phi_2(y)$
&
&
$\Phi_2(y)\star\Phi_1(x)$
&
&
$\Phi_2(y)\star\Phi_1(y)$\makebox[3\unitlength]{}\\
\ \\
\multicolumn{7}{c}{
Fig. 1 \ :\ Cyclic property on the boundary of a disc
}
\end{tabular}
\setlength{\unitlength}{\backu}
\vspace*{1.0cm} 

Similarly the three point function 
\begin{equation}
I_3[\Phi_1,\Phi_2,\Phi_3] \equiv \int d^Dx \ \omega(x) \left. (\Phi_1(x) 
\bm{\star} \Phi_2(x) \bm{\star} \Phi_3(x)) \right|_{y=x} 
\end{equation}
is given by 
\begin{eqnarray}
I_3[\Phi_1,\Phi_2,\Phi_3] 
&=& \int d^Dx \{ \ \omega(x) \cdot \Phi_1(x) * \Phi_2(x) * \Phi_3(x)
 -\frac{i}{6} (G^{\mu \lambda}G^{\nu \rho}-\theta^{\mu \lambda}
\theta^{\nu \rho}) H_{\lambda \rho \sigma} \  x^{\sigma} \nonumber \\
&& \qquad \times ( \partial_{\mu}\Phi_1 * \partial_{\nu}\Phi_2 * \Phi_3 +
\partial_{\mu}\Phi_1 * \Phi_2 * \partial_{\nu}\Phi_3  + \Phi_1 
*\partial_{\mu}\Phi_2 * \partial_{\nu}\Phi_3) \nonumber \\
&& \qquad + \frac{1}{18} (G^{\mu \lambda}\theta^{\nu \alpha}-
\theta^{\mu \lambda}G^{\nu \alpha}) \  G^{\sigma \rho} \ 
H_{\lambda \rho \alpha} \ 
\partial_{\mu}\Phi_1 * \partial_{\nu}\Phi_2 * \partial_{\sigma}\Phi_3 \  \}. 
\label{cycl3}
\end{eqnarray}
In the $\alpha' \rightarrow 0$ limit those terms in (\ref{I2}) and 
(\ref{cycl3}) which contain the metric $G^{\mu \nu}$ drop and agree with the 
results of correlation functions obtained in \cite{CS}.

It is not difficult to show that the three point function (\ref{cycl3})  
satisfies 
\begin{equation}
I_3[\Phi_1,\Phi_2,\Phi_3] =I_3[\Phi_3,\Phi_1,\Phi_2]= I_3[\Phi_2,\Phi_3,
\Phi_1].
\label{cyc3}
\end{equation}
We can also show the following identities. 
\begin{eqnarray}
&& \int d^Dx \ \omega \left. (\Phi_1(x) \bm{\star} \Phi_2(x) \bm{\star} 
\Phi_3(x)) \right|_{y=x} = \int d^Dx \ \omega \left. (\Phi_3(y) \bm{\star} 
\Phi_2(y) \bm{\star} 
\Phi_1(y) ) \right|_{y=x} \nonumber \\
&=& \int d^Dx \ \omega \left. (\Phi_1(x) \bm{\star} \Phi_2(x) \bm{\star} 
\Phi_3(y)) \right|_{y=x} = \int d^Dx \ \omega \left. (\Phi_1(x) \bm{\star} 
\Phi_3(y)\bm{\star} \Phi_2(x) ) \right|_{y=x} \nonumber \\
&=& \int d^Dx \ \omega \left. (\Phi_3(y) \bm{\star} \Phi_1(x) \bm{\star} 
\Phi_2(x)) \right|_{y=x} = \int d^Dx \ \omega \left. (\Phi_1(x) \bm{\star} 
\Phi_3(y)\bm{\star} \Phi_2(y) ) \right|_{y=x} \nonumber \\
&=& \int d^Dx \ \omega \left. (\Phi_3(y) \bm{\star} \Phi_1(x) \bm{\star} 
\Phi_2(y)) \right|_{y=x} = \int d^Dx \ \omega \left. (\Phi_3(y) \bm{\star} 
\Phi_2(y)\bm{\star} \Phi_1(x) ) \right|_{y=x}
\label{cy3}
\end{eqnarray}
The locations of the fields can be shifted from one end $x$ of the 
string to the other $y$, and {\em vice versa}, as long as the ordering on a 
circle is unchanged.  This is natural from the string theory
point of view.

Higher point functions $I_n[\Phi_1, \cdots,\Phi_n]$ do not have the 
cyclic property. In \cite{CS} it was shown that in order to obtain the four 
point function in terms of the non-associative product $\bullet$ one must 
take a specific linear combination of the ordered products of the functions 
with various positionings of the parenthesis. In our approach the correlation 
functions will be obtained by functional derivatives of the effective action 
and various orderings will be automatically taken into account due to the 
bose symmetry.  

The cyclic property (\ref{cyc2}), (\ref{cycl2}), (\ref{cyc3}), (\ref{cy3}),
and the agreement of (\ref{I2}), (\ref{cycl3}) with the results of \cite{CS} 
in the $\alpha' \rightarrow 0$ 
limit justifies the prescription (\ref{reduced}).

Let us now turn to the gauge theory action (\ref{gauge}).  
The reduced effective action which corresponds to this is given by 
\begin{eqnarray}
&& S_{\rm{reduced} \ \rm{effective} \ \rm{action}} [ A_{\mu}, \ 
\overline{A}_{\mu}, \  X^m, \ \overline{X}^m ] \ = \ \int d^Dx \sqrt{G} \ 
\omega(x)\ [ \ {\cal L}(x,y) \ ]|_{y=x} \nonumber \\
\qquad \qquad &&=  -\frac{1}{4g^2_{\rm{YM}}}\int d^Dx \ \sqrt{G} \ \omega(x) \ 
\cdot  \{ \ G^{\mu \nu} G^{\lambda \rho} \  [D_{\mu}, \  
D_{\lambda} ]_{\star} \ \bm{\star} \  [D_{\nu}, \ D_{\rho} ]_{\star} 
\nonumber \\ 
&& \qquad \qquad \qquad    - 2 G^{\mu \nu} \  [D_{\mu}, \  X^m ]_{\star} 
\bm{\star} [D_{\nu}, \ X^m ]_{\star}  + [X^m, \ X^n]_{\star}^2 \nonumber \\
&& \qquad \qquad \qquad +\ G^{\mu \nu} G^{\lambda \rho} \  
[\overline{D}_{\mu}, \ \overline{D}_{\lambda} ]_{\star} \ \bm{\star} \  
[\overline{D}_{\nu}, \ \overline{D}_{\rho} ]_{\star} \nonumber \\ 
&& \qquad \qquad \qquad  \left. - 2 G^{\mu \nu} \ [\overline{D}_{\mu}, \  
\overline{X}^m ]_{\star} \bm{\star} [\overline{D}_{\nu}, \ 
\overline{X}^m ]_{\star}  + [\overline{X}^m, \ 
\overline{X}^n]_{\star}^2 \} \right|_{y=x}, 
\label{red}
\end{eqnarray}
where ${\cal L}(x,y)$ is the Lagrangian density defined in (\ref{gauge}).
The restriction $y=x$ here breaks some part of the gauge symmetry. The reduced
effective  action (\ref{red}), however, turned out to be still invariant 
under the gauge transformation with the gauge function 
\begin{equation}
\Lambda(x,y)= \hat{\Lambda}(x)-\hat{\Lambda}(y). 
\label{residual gauge}
\end{equation}
This follows from the identity
\begin{equation}
\int d^Dx \ \omega(x) \left. \left\{ [\hat{\Lambda}(x)- \hat{\Lambda}(y), 
f(x,y) ]_{\star} \right\} \right|_{y=x} =0,
\label{identity}
\end{equation}
which can be proved by expanding $f(x,y)$ as $\sum_n g_n(x) \bm{\star} h_n(y)$
and using the identities (\ref{cy3}). Because {\boldmath $\bm{\star}$} 
contains derivatives, the integrand of (\ref{identity}) does not 
trivially vanish, even if we set $y=x$.

Let us note that in (\ref{red}) the product inside the bracket is  
{\boldmath $\bm{\star}$}. Hence the reduced action is not an ordinary action 
in the $D$-dimensional space, because the product in the integrand contains 
both derivatives $\partial / \partial x$, $ \ \partial / \partial y$.
The gauge transformation is also defined in the $(x,y)$ space. 
To write down the gauge invariant action in the ordinary sense we must 
introduce a set of coordinates $(x,y)$ and consider 2$D$  dimensional action
integral, as we did in the previous section.

Finally, in order to derive the correlation functions we must further set 
$A_{\mu}=\hat{A}_{\mu}(x)$, $\overline{A}_{\mu}=\hat{A}_{\mu}(y)$, 
$X^m=\hat{X}^m(x)$ and $\overline{X}^m=\hat{X}^m(y)$ before taking the 
functional derivatives of (\ref{red}), because these describe the gauge 
particles emitted and/or absorbed from the $\sigma=0$ and $\pi$ ends of the 
open string and the fluctuations at the two ends, respectively. The same 
functions at both ends because of the bose symmetry. Then this reduced action 
becomes a functional of $\hat{A}_{\mu}$, $\hat{X}^m$.

We must mention some ambiguity in our prescription. When we perform 
a partial integration or use the cyclic property (\ref{cyclicproperty}) in 
(\ref{gauge}), we will obtain different integrands which will, however, lead 
to the same result. The corresponding reduced integral does not have this 
property. This is unavoidable, because when we reduce the integral, we insert 
into the integrand a function proportional to a delta function. The result 
will depend on where this delta function is inserted and we must specify this. 
Our proporsal is to reduce the integral in the standard form of the action 
(\ref{gauge}).

Because we have not determined ${S^{\mu \nu}}_{\lambda}$, 
${T^{\mu \nu}}_{\lambda}$, ${U^{\mu \nu}}_{\lambda}$ completely, we cannot
decide whether (\ref{red}) reproduces the open string amplitudes in the low 
energy limit. Moreover such an analysis will meet difficulties, 
because one cannot generally define free states in an asymptotically non-flat 
space. We, however, expect that once ${S^{\mu \nu}}_{\lambda}$, 
${T^{\mu \nu}}_{\lambda}$, ${U^{\mu \nu}}_{\lambda}$ are determined, the gauge 
invariance is strong enough to restrict the form of the effective action.

\section{Discussions}
\hspace{5mm}
In this paper we considered an open string theory in an NS-NS $B$ field 
background with a nonzero constant field strength $H=dB$. The background 
space is curved although if $H$ is small, the metric $g_{\mu \nu}$ is 
constant up to ${\cal O}(H^1)$. We performed a perturbative analysis of the 
commutation relations of the string coordinates without oscillators up to 
${\cal O}(H^1)$ and found that the coordinate $x$ of one  end of the string 
does not commute with the coordinate $y$ of the other. We then constructed an 
associative and noncommutative product {\boldmath $\star$} which realizes the 
commutation relations of the coordinates. This product is an operation for 
functions of both the coordinates $x$, $y$ and even if two functions of $x$ 
only are multiplied, the result is a function of both $x$ and $y$. In this 
way we are lead to consider bi-local fields which depends on the coordinates 
of both ends of the string. 

We then found that derivatives $\partial/\partial x^{\mu}$, 
$\partial/\partial y^{\mu}$ do not satisfy Leibnitz rule with the 
{\boldmath $\star$} product and we are forced to modify the derivatives. 
At present the modified derivatives are not uniquely determined, but 
remarkably, the new derivatives $\nabla/\nabla x^{\mu}$, $\nabla/\nabla 
y^{\mu}$ turned out to be rewritten as $\partial/\partial x^{\prime \mu}$, 
$\partial/\partial y^{\prime \mu}$, respectively, by some `coordinate 
transformation' $x,\ y \rightarrow x', \ y'$. 
The commutation relations of the primed variables coincide with those of 
the unprimed ones with $H=0$. Hence if we combine the $x$ and $y$ coordinates 
and consider 2$D$  dimensional space $(x,y)$, the primed coordinate system
appears to be flat. This provides us with a clue to write down the gauge 
theory action. The open string metric $\hat{G}_{\mu \nu}(x) = (g \{1-(g^{-1} 
B(x))^2 \})_{\mu \nu}$ is curved and it may seem difficult to write down the  
action without spoiling the gauge invariance due to the coordinate dependent 
metric. Our proporsal is to write down the action in the flat $(x',y')$ frame 
and then transform it to the $(x,y)$ frame. In this way we obtained an action 
integral for the noncommutative gauge theory in curved backgrounds. This 
gauge theory lives in a double-dimensional space, {\it i.e.}, the space with 
coordinates 
$(x^{\mu}, y^{\mu}), \ \mu=1,2, \ldots, D$. This action is invariant under 
gauge transformations with $x$, $y$ dependent gauge parameters. We then 
took a limit $y \rightarrow x$ and proposed a prescription to reduce the 
action integral to that in $D$-dimensional space with a coordinate $x$. 
This reduced action still has a part of the gauge symmetry. This will provide 
a `local' low energy effective action for string amplitudes. 

Let us now turn to the application of our results to the Matrix model. 
The formulation of the Matrix model for curved backgrounds has been 
investigated \cite{MD} but is not yet well established. As mentioned above, 
the backgound space is curved for nonvanishing $H$. The closed string metric 
$g_{\mu \nu}$ is flat only up to ${\cal O}(H^1)$ and the open string metric 
$\hat{G}_{\mu \nu}(x)$ is curved already at ${\cal O}(H^1)$\cite{CS}. On the 
other hand in the action (\ref{gauge}) the metric $G_{\mu \nu}$ is flat and 
has the ordinary form of the gauge theory action in the flat space. This is 
based on the fact that we can deform the algebra of the coordinates $x$, $y$ 
to that of the flat background by some  coordinate transformation $(x,y) 
\rightarrow (x',y')$. Although we do not have a proof that this persists to 
all orders in $H$, let us assume this here. Then this observation suggests 
the following general formulation of the Matrix model for curved 
backgrounds.   
\begin{itemize}
\item Double the number of the matrices of the ordinary matrix model.
Denote these as $X^{\mu}$ and $\overline{X}^{\mu}$. This is natural because 
the open string has two ends. In this paper we considered only the bosonic 
string.  In the case of the superstring the fermion fields on the D-brane 
must be also doubled. 

\item The action of the Matrix model is given by
\begin{eqnarray}
S_{\rm{Matrix}} &=& \frac{1}{4 g_{\rm{YM}}^2} \int dt \  Tr \{ \ 2
(D_0 \ X^m)^2 + 2(\overline{D}_0 \ \overline{X}^m)^2 +\ [X^{m}, \ X^{n}]^2 \  
 +[\overline{X}^{m}, \ \overline{X}^{n}]^2  \nonumber \\
&& \qquad \qquad \qquad \qquad  + \ \mbox{fermions} \ \}.
\label{Matrix}
\end{eqnarray}
\item 
The action (\ref{Matrix}) is background independent.  The space-time metric 
is Minkowskian. We can introduce 
backgrounds as follows. By a separate linear coordinate transformation of 
$X^{\mu}$, $\overline{X}^{\mu}$ the constant metric $G_{\mu \nu}$ can be put 
in. The matrix multiplication is associative and noncommutative and there 
should be an isomorphism between matrices $X^{\mu}$, $\overline{X}^{\mu}$ and 
the bi-local fields whose multiplication rule obeys the product {\boldmath 
$\star$}. By the replacement $X^{\mu} \rightarrow iD_{\mu}$, 
$\overline{X}^{\mu} \rightarrow i \overline{D}_{\mu}$ $(\mu=1,2, \ldots, D)$ 
and $Tr \rightarrow \int d^Dx d^Dy V^{-1} \ G \ \Omega(x,y)$, we will obtain 
the action (\ref{gauge}). The product {\boldmath $\star$} and the gauge 
covariant derivatives $D_{\mu}$, $\overline{D}_{\mu}$ will introduce the 
background dependence. 
\end{itemize}
 
For the above formulation to be valid we must obtain the classical 
solution for the Matrix model action (\ref{Matrix}) and then find a way to 
systematically determine the star product {\boldmath $\star$} and the 
covariant derivatives $D$, $\overline{D}$ and demonstrate that the gauge 
theory action (\ref{gauge}) can be obtained. We hope to report on the analysis
in the future.




\end{document}